# Extraction of 3D field maps of magnetic multipoles from 2D surface measurements with applications to the optics calculations of the large-acceptance superconducting fragment separator BigRIPS

Hiroyuki Takeda [a],*, Toshiyuki Kubo [a], Kensuke Kusaka [a], Hiroshi Suzuki [a,*], Naohito Inabe [a], Jerry A. Nolen [b]

[a] *RIKEN Nishina Center, RIKEN, 2-1 Hirosawa, Wako, Saitama 351-0198, Japan*

[b] *Argonne National Laboratory (ANL), 9700 S. Cass Avenue, Argonne, Illinois 60439, USA*

**Abstract**

The fringing fields of magnets with large apertures and short lengths greatly affect ion-optical calculations. In particular, for a high magnetic field where the iron core becomes saturated, the effective lengths and shapes of the field distribution must be considered because they change with the excitation current. Precise measurement of the three-dimensional magnetic fields and the correct application of parameters in the ion-optical calculations are necessary. First we present a practical numerical method of extracting full 3D magnetic field maps of magnetic multipoles from 2D field measurements of the surface of a cylinder. Using this novel method we extracted the distributions along the beam axis for the coefficient of the first-order quadrupole component, which is the leading term of the quadrupole components in the multipole expansion of magnetic fields and proportional to the distance from the axis. Higher order components of the 3D magnetic field can be extracted from the leading term via recursion relations. The measurements were done for many excitation current values for the large-aperture superconducting triplet quadrupole magnets (STQs) in the BigRIPS fragment separator at the RIKEN Nishina Center RI Beam Factory. These distributions were parameterized using the Enge functions to fit the fringe field shapes at all excitation current values, so that unmeasured values are interpolated. The extracted distributions depend only on the position along the beam axis, and thus the measured three-dimensional field can easily be parameterized for ion-optical calculations. We implemented these parameters in the ion-optical calculation code COSY INFINITY and realized a first-order calculation that incorporates the effect of large and varying fringe fields more accurately. We applied the calculation to determine the excitation current settings of the STQs to realize various optics modes of BigRIPS and the effectiveness of this approach has been demonstrated.

*Corresponding authors:
takeda@ribf.riken.jp (Hiroyuki Takeda) and hsuzuki@ribf.riken.jp (Hiroshi Suzuki)

## 1. Introduction

Magnetic devices, including bending magnets and quadrupole magnets, are widely used in accelerators, beam transport systems, and spectrometers. To design and analyze ion-optical systems that are built using these magnetic devices, various codes have been developed. They are grouped into two main categories.

The first category calculates a transfer matrix connecting phase space variables at the initial and final points order-by-order based on a Taylor expansion [1]. The physical meanings of the first- and low-order matrix elements are relatively easy to understand, and the ion-optical design can be achieved easily. Rough ion-optics are determined at the first order and are corrected order-by-order by calculating the high-order aberrations. Thanks to this convenience, this type of calculation has been widely used to design beam transport systems. In particular for magnets for which a sharp cut-off approximation is effective, many easy-to-use analytic codes [2-5] exist for ion-optical design and analysis.

The other category is the ray-trace type, which solves an equation of motion and tracks the path of individual particles [6, 7] using a three-dimensional field distribution that is measured or calculated beforehand. With this method, it is possible to calculate orbits even for a spectrometer with a complex magnetic field distribution. However, since the amount of calculation is large and it is difficult to extract the ion-optical information directly, it is not easy to use for the design and analysis of a beam transport system.

The use of large-aperture high-field magnets has been increasing recently because of the requirements of the secondary beam transport system, especially for radioactive isotope (RI) beams, to provide spread beams both in angles and momenta with high magnetic rigidity. RI Beam Factory (RIBF) [8] in RIKEN is a typical example. In order to study the properties and reactions of exotic nuclei far from the stability line, including searches for new isotopes, the RIBF provides high-intensity unstable nuclear beams as secondary beams by means of in-flight fission [9] or projectile fragmentation [10], which occur by impinging heavy ion beams, including uranium beams, on a target. BigRIPS [11-13] is an in-flight RI beam separator used to separate RIs and produce and transport RI beams. It is characterized by its large acceptance, large magnetic rigidity and the adoption of a two-stage separator scheme. In order to deliver beams having wide angles and momenta and neutron-rich nuclei with high magnetic rigidity produced by in-flight fission of uranium, superconducting triplet quadrupole magnets (STQs) [14] are adopted, which enable the achievement of large-aperture and strong magnetic fields. It realizes an angular acceptance of ±40 mrad (horizontal) and ±50 mrad (vertical), and a momentum acceptance of ±3%. Furthermore not only are the apertures of the quadrupole magnets large, but their lengths must be short in order to achieve such large acceptances. Because RIs must be provided with particle identification for each event [15], the BigRIPS is also designed as a spectrometer with a high momentum resolution in order to achieve high particle identification power. RIs are generated with a wide range in the magnetic rigidity from 2 Tm to 9.7 Tm according to experimental conditions and requirements. Thus the precise setting of the magnets covering such a wide range is indispensable to realize the designed performance of the spectrometer.

In the case of large-aperture and short-length magnets with strong magnetic fields, the fringing field is generally very large, and the shape and effective length of the magnetic field distribution change with the excitation current due to the saturation of the iron core. Although the first- and higher-order transfer matrices of the fringing fields can be taken into account analytically in terms of fringing field integrals [16] in many ion-optical codes [2-5], they are not so convenient to be applied in our situation where the fringing fields vary drastically according to the excitation current. An ion-optical calculation that takes into account the effect of the varying fringing field is required to realize precise setting of the large-aperture, short-length and high-field magnets. We use the code COSY INFINITY [17] because it can calculate the effect of almost any shape of fringing field which can be parameterized with excitation current while maintaining the utility of a transfer map type ion-optical simulation.

Another issue is the decomposition of the so-called higher-order pseudo terms of the measured three-dimensional magnetic field. For instance, the measured magnetic field of quadrupole magnets generally includes high-order pseudo components proportional to $r^3, r^5, \cdots$, where $r$ is the distance from the axis, in addition to the leading term of the quadrupole component proportional to $r$. These pseudo components originate from the changes of the magnetic field in the direction along the beam axis, and do not have to be taken into account in many cases when small-aperture and long magnets are used and the magnetic field distribution along the beam axis is almost uniform. In our situation where the large-aperture and short-length STQs are used, it is indispensable to correctly extract the higher-order pseudo quadrupole components as well as the first-order quadrupole components from the measured magnetic field maps even for first-order ion-optical simulations. At first glance, it appears that field map data measured at different radii are necessary to solve the $r$ dependence. Indeed, in an early attempt of such high-order decomposition performed by G. R. Moloney *et al.* [18], a polynomial in $r$ was assumed to extract the $r$ dependence by fitting field maps measured at different radii. On the other hand, theorems have been developed to show that 3D magnetic fields interior to a closed 2D surface can be extracted from magnetic field maps on that surface. A discussion and some applications of these theorems are presented, for example, in Refs. [19, 20].

We present below a new practical numerical procedure to solve the problem. With this method it is possible to extract full 3D magnetic field maps of a multipole magnet in cylindrical coordinates using only 2D data measured at one radius for one component on the surface of a cylinder that passes through the multipole. The extracted leading term of the quadrupole component (i.e. the leading term of the $\cos(2\theta)$ multipole term) is expressed as a function that depends only on the position $z$ along the beam axis, and all the other high-order pseudo components of the 3D magnetic field, which also have a $\cos(2\theta)$ angular dependence but higher radial power dependence, can be derived from this function via recursion relations. Similarly, higher order multipole fields can be extracted if necessary. Therefore, by measuring the magnetic field distributions in the cylindrical coordinates at small excitation current steps, a function for the quadrupole components at each excitation current, as well as the full three-dimensional field distributions, can be obtained. In addition, because the function derived in this method depends only on $z$, it is relatively easy to parameterize it by using, for example, field-dependent parameters of the Enge function [21]. This method makes it convenient to determine ion-optical simulations of tunes for BigRIPS for operation at arbitrary rigidity settings. This method can be applied to any magnet having rotational multipole symmetry about a straight axis. We have used this method for the superconducting triplet quadrupole magnets (STQs), which are commonly used in the BigRIPS.

In this paper, we present a cylindrical coordinate description and general equation for multipole decomposition and describe the steps for extracting the leading terms from the magnetic field measurement data. We emphasize that the measurement data for one radius of one component in the cylindrical coordinates, i.e. 2D field measurements on the surface of a cylinder, are sufficient to determine the full three-dimensional magnetic field in the whole region under consideration. We use our BigRIPS separator and STQ magnets as examples to show how to analyze field measurement results and to extract the leading term of the quadrupole component using the above method, and to describe fitting by the Enge function and excitation current interpolation of the coefficients. Ion-optical calculations of the BigRIPS with the obtained quadrupole fields are also described.

## 2. Formalism

### 2.1 Recursion relations for components of the 3D magnetic field

Quadrupole, sextupole, and other higher-order multipole magnets have a straight beam axis and a certain rotational symmetry around the axis. Cylindrical coordinates are most suited to describing their magnetic fields. In the following discussion, we use cylindrical coordinates where $z$ denotes the beam axis, $r$ denotes the distance from the beam axis, and $\theta$ denotes the azimuthal angle from a horizontal plane. When there is no time dependency and in a region with no current, Maxwell's equations for the magnetic field $\vec{B}$ are simplified as $\vec{\nabla}\cdot\vec{B}=\vec{\nabla}\times\vec{B}=0$. By introducing the magnetic scalar potential $\Phi$, where $\vec{B}=\vec{\nabla}\Phi$, the problem is reduced to solving Laplace's equation, $\vec{\nabla}^2\Phi=0$. The fundamental solutions of the Laplace's equation in cylindrical coordinates are widely known to be the Bessel functions $J_n(\kappa r), N_n(\kappa r)$, trigonometric functions $\cos n\theta, \sin n\theta$, and exponential functions $\exp(\pm\kappa z)$ for $r$, $\theta$, and $z$, respectively, and their linear combination is a general solution. Here, $n$ and $\kappa$ are constants introduced in the process of separation of variables and are arbitrary values in principle. However, if $\Phi$ is required to be a single-valued function, then $n$ must be an integer. The solution when $n$ = 0 represents a solenoidal magnetic field independent of the azimuthal angle $\theta$, and because it is not dealt with here, $n \geq 1$ is assumed hereafter. When $\kappa =$ 0, $\Phi$ will be a function independent of $z$. Realistic magnets have finite lengths, and the magnetic fields vary with $z$. Therefore, a solution for $\kappa$ = 0 is not applicable in this case, and so we assume $\kappa \neq 0$. For convenience in the following discussions, $\kappa^{-1}$ is replaced with $r_0$. Out of the two kinds of Bessel functions $J_n(\kappa r)$ and $N_n(\kappa r)$, only $J_n(\kappa r)$ are selected as the solutions for the $r$ coordinate, while $N_n(\kappa r)$ are not applicable because they diverge at $r$ = 0. When one $J_n(\kappa r)$ is expanded as a power series of $r$, the lowest order term is $r^n$, and it continues as $r^{n+2}, r^{n+4}, \cdots, r^{n+2m}, \cdots$, for each $n$.

Therefore, the solution satisfying our boundary conditions can be written as follows.

$$\Phi(r,\theta,z) = \sum_{n=1}^{\infty}\sum_{m=0}^{\infty}\left[\frac{b_{n,m}(z)r_0}{n+2m}\left(\frac{r}{r_0}\right)^{n+2m}\sin n\theta + \frac{a_{n,m}(z)r_0}{n+2m}\left(\frac{r}{r_0}\right)^{n+2m}\cos n\theta\right]. \tag{1}$$

A similar expression was presented, for example, in Refs. [18, 22]. The linear combination of $\exp(\pm\kappa z)$ are denoted as $b_{n,m}(z)$ and $a_{n,m}(z)$. Both of them have the same physical dimension representing magnetic fields for any $n$ and $m$. $b_{n,m}(z)$ and $a_{n,m}(z)$ are designated as normal and skew components, respectively, because $a_{n,m}(z)$ terms vanish if mid-plane symmetry exists. In realistic magnetic field measurements, it is difficult to align the angle of the field mapping device or Hall sensors exactly with the symmetry plane of a magnet, and this effect can generate the apparent skew components $a_{n,m}(z)$. Such measurement errors can be removed by retaking the origin of $\theta$ per Hall sensor. This assumes that real skew components arising from magnet coil misalignments or errors are negligible. In this paper, we discuss primarily the normal components only assuming the mid-plane symmetry of the magnet. The scale radius $r_0$ can be selected arbitrarily. It is convenient to set it to a pole tip radius, a warm bore radius or a radius used to measure the magnetic field map. When the scale radius is changed from $r_0$ to $r_0'$, $b_{n,m}(z)$ is simply scaled by $(r_0'/r_0)^{n+2m-1}$. It is worthy to note that the leading term $b_{n,0}$ is replaced by $(r_0'/r_0)^{n-1}b_{n,0}$. This relationship is useful when comparing the results of magnetic field measurement at different radii. Substituting equation (1) in the Laplace's equation $\vec{\nabla}^2\Phi = 0$ yields the following recurrence relation.

$$b_{n,m}(z) = -\frac{r_0^2}{4m(n+m)}\cdot\frac{n+2m}{n+2(m-1)}\cdot\frac{d^2 b_{n,m-1}(z)}{dz^2}\quad(m \geq 1). \tag{2}$$

Once we know the $b_{n,0}(z)$ distribution, which is a function of $z$ only, all $b_{n,m}(z)$ values can be calculated from this equation, and the complete three-dimensional magnetic field map for the region in consideration can be obtained. By taking gradient of $\Phi$, the magnetic field in cylindrical coordinates can be described as follows:

$$\begin{cases} B_r(r,\theta,z) = \sum_{n=1}^{\infty} B_{r,n}(r,z)\sin n\theta\,, \\ B_\theta(r,\theta,z) = \sum_{n=1}^{\infty} B_{\theta,n}(r,z)\cos n\theta\,, \\ B_z(r,\theta,z) = \sum_{n=1}^{\infty} B_{z,n}(r,z)\sin n\theta\,, \end{cases}$$

with

$$\begin{cases} B_{r,n}(r,z) \equiv \left(\frac{r}{r_0}\right)^{n-1} \sum_{m=0}^{\infty} b_{n,m}(z)\left(\frac{r}{r_0}\right)^{2m}, & (3a) \\ B_{\theta,n}(r,z) \equiv \left(\frac{r}{r_0}\right)^{n-1} \sum_{m=0}^{\infty} \frac{n}{n+2m} b_{n,m}(z)\left(\frac{r}{r_0}\right)^{2m}, & (3b) \\ B_{z,n}(r,z) \equiv \left(\frac{r}{r_0}\right)^{n} \sum_{m=0}^{\infty} \frac{r_0}{n+2m}\frac{db_{n,m}(z)}{dz}\left(\frac{r}{r_0}\right)^{2m}. & (3c) \end{cases}$$

Here, $B_r$, $B_\theta$, and $B_z$ represent the magnetic field components in the radial direction, azimuthal angle direction, and beam axis direction, respectively. The number $n$ corresponds to the multipole order ($n$ = 1 represents a dipole magnet, $n$ = 2 represents a quadrupole magnet, etc.).

### 2.2 Numerical method for the extraction of full 3D B-field maps from 2D B-field components measured on the surface of a cylinder of fixed radius

The multipole components $B_{r,n}$, $B_{\theta,n}$, and $B_{z,n}$ can be obtained by standard Fourier analysis, such as $B_{r,n}(r,z) = \int_0^{2\pi} B_r(r,\theta,z)\sin n\theta\, d\theta$, from the measured magnetic field maps $B_r$, $B_\theta$, and $B_z$. In addition to the leading term $(m = 0)$ proportional to $r^{n-1}$, pseudo-multipole terms $(m \geq 1)$ also appear. Pseudo-multipole terms have the angular dependence of the leading term, such as $\cos(2\theta)$ for a quadrupole, but have a higher order radial dependence, such as $r^3$ rather than $r$ for a quadrupole. As shown in equation (2), these pseudo-multipole terms are derived from the $z$ derivative of the leading term $b_{n,0}$. Therefore, these terms are small for sufficiently long magnets for which the magnetic field distributions are considered to be constant, and are often ignored in conventional analysis. Also, note that since the pseudo terms are higher even derivatives of the leading terms, they will mathematically integrate to zero through the magnet, which also minimizes their effect on the ion optics. However, they cannot be ignored for large-aperture short-length magnets such as the STQs used in the BigRIPS. Therefore, separating each term is important. Because all the pseudo-multipole terms can be calculated from $b_{n,0}$, as mentioned above, the problem is reduced to extracting $b_{n,0}$.

$b_{n,0}$ can be extracted as follows. First, consider equation (2). In order to avoid differential operation of high orders, apply the Fourier transform

$$\tilde{b}_{n,m}(k) = \int_{-\infty}^{\infty} b_{n,m}(z) e^{-ikz} dz$$

to equation (2):

$$\begin{aligned}\tilde{b}_{n,m}(k) \quad &= -\frac{r_0^2}{4m(n+m)} \cdot \frac{n+2m}{n+2(m-1)} (-ik)^2 \tilde{b}_{n,m-1}(k) \\ &= q_m \tilde{b}_{n,m-1}(k).\end{aligned}$$

Here, the coefficient

$$q_m \equiv \frac{(r_0 k)^2}{4m(n+m)} \cdot \frac{n+2m}{n+2(m-1)}$$

is a non-dimensional number. By applying this recursion relation repeatedly, $\tilde{b}_{n,m}$ can be written in terms of $\tilde{b}_{n,0}$ as follows:

$$\begin{aligned}\tilde{b}_{n,m}(k) \quad &= q_m \tilde{b}_{n,m-1}(k) \\ &= q_m q_{m-1} \tilde{b}_{n,m-2}(k) \\ &\vdots \\ &= q_m q_{m-1} \cdots q_1 \tilde{b}_{n,0}(k) \\ &\equiv p_m \tilde{b}_{n,0}(k),\end{aligned}$$

where

$$p_m \equiv q_m q_{m-1} \cdots q_1.$$

The derivation of the above equation is valid only when $m \geq 1$, but the final equation

$$\tilde{b}_{n,m}(k) = p_m \tilde{b}_{n,0}(k), \tag{4}$$

is also valid for $m = 0$ when we define $p_0 = 1$. Note that these recursion relations of the Fourier transforms do not require the evaluation of numerical derivatives.

Using this result, the Fourier transform of (3a) becomes

$$\begin{aligned}\tilde{B}_{r,n}(k) \quad &= \sum_{m=0}^{\infty} \left(\frac{r}{r_0}\right)^{n+2m-1} \tilde{b}_{n,m}(k) \\ &= \sum_{m=0}^{\infty} \left(\frac{r}{r_0}\right)^{n+2m-1} p_m \tilde{b}_{n,0}(k).\end{aligned}$$

Thus,

$$\tilde{b}_{n,0}(k) = \tilde{B}_{r,n}(k) \Big/ \sum_{m=0}^{\infty} \left(\frac{r}{r_0}\right)^{n+2m-1} p_m. \tag{5}$$

$\tilde{B}_{r,n}(k)$ itself can be obtained from the Fourier transform of the measured magnetic field map as

$$\tilde{B}_{r,n}(k) = \int_{-\infty}^{\infty} B_{r,n}(r,z) e^{-ikz} dz.$$

Thus the right side of the equation (5) only shows the quantities obtained from the measured magnetic field map data, which means that $\tilde{b}_{n,0}(k)$ was successfully extracted from the measured data. The Fourier transform of (3b) provides a similar result for $B_{\theta,n}$,

$$\tilde{b}_{n,0}(k) = \tilde{B}_{\theta,n}(k) \Big/ \sum_{m=0}^{\infty} \left(\frac{r}{r_0}\right)^{n+2m-1} \frac{np_m}{n+2m}. \tag{6}$$

Obviously it is useful when the scale radius $r_0$ is chosen to be the radius at which the field map measurements are performed. In this case, equations (5) and (6) become more convenient forms:

$$\tilde{b}_{n,0}(k) = \tilde{B}_{r,n}(k) \Big/ \sum_{m=0}^{\infty} p_m, \tag{5'}$$

and

$$\tilde{b}_{n,0}(k) = \tilde{B}_{\theta,n}(k) \Big/ \sum_{m=0}^{\infty} \frac{np_m}{n+2m}. \tag{6'}$$

Both $B_{r,n}$ and $B_{\theta,n}$ can be used to extract $\tilde{b}_{n,0}(k)$ independently. When $B_{\theta,n}$ is used, the larger $m$ is, the less important the term becomes due to the reduction factor $n/(n+2m)$. Therefore, $B_{\theta,n}$ converges faster than $B_{r,n}$, and the error is expected to be relatively small.

Mathematically, $B_{z,n}$ data can be used to extract $\tilde{b}_{n,0}(k)$ by the similar procedure. However, $B_{z,n}$ is originally small in magnitude compared to $B_{r,n}$ or $B_{\theta,n}$ and has a larger relative measurement error, and hence we did not use $B_{z,n}$ data to extract $\tilde{b}_{n,0}(k)$.

Finally, by performing the inverse Fourier transform of the extracted $\tilde{b}_{n,0}(k)$, $b_{n,0}(z)$ is obtained.

$$b_{n,0}(z) = \frac{1}{2\pi} \int_{-\infty}^{\infty} \tilde{b}_{n,0}(k) e^{+ikz} dk.$$

Once the $b_{n,0}(z)$ distribution is obtained, any pseudo-multipole term $b_{n,m\geq 1}(z)$ can be obtained by the recurrence relation (2), and a complete three-dimensional magnetic field map that satisfies the Maxwell's equations is obtained automatically. This fact is very useful when modeling a three-dimensional magnetic field map with a simple mathematical expression. Another way to extract the pseudo multipole terms is to use the relation (4). The inverse Fourier transform yields

$$\begin{aligned} b_{n,m}(z) &= \frac{1}{2\pi} \int_{-\infty}^{\infty} \tilde{b}_{n,m}(k) e^{+ikz} dk \\ &= \frac{1}{2\pi} \int_{-\infty}^{\infty} p_m \tilde{b}_{n,0}(k) e^{+ikz} dk, \end{aligned}$$

in which any numerical derivatives are not involved.

We emphasize that the advantage of extracting $b_{n,0}(z)$ is that the measurement result for any one of the magnetic field components in cylindrical coordinates $(B_r, B_\theta, B_z)$ at any radius is sufficient in principle. In addition, in the process of extracting $b_{n,0}(z)$, the random errors included in the measured results are averaged during the integration process of the Fourier transform, and the errors are expected to become small.

The steps of the process described in this section are summarized as a diagram in Fig. 1.

## 3. Application to multipole analysis of STQ

Here, the procedure described in the previous section is applied to the STQs used in the BigRIPS, which consists of six room-temperature dipole magnets (D1–D6) and 14 STQs (STQ1–14). STQs of the same design (STQ15–25) are also used for the ZeroDegree spectrometer [11-13] and the RI-beam delivery lines connected to the large-acceptance multi-particle spectrometer SAMURAI [23] and high-resolution SHARAQ spectrometer [24], which are located downstream of BigRIPS.

STQ1, the most upstream, is an air-core type, and the rest are superferric type. Each STQ is configured as either Q500-Q800-Q500 or Q500-Q1000-Q500. Here, Q500, Q800, and Q1000 are superconducting quadrupole magnets with four-fold symmetry, which have magnet lengths of 500 mm, 800 mm, and 1000 mm, respectively. A superconducting sextupole coil is superimposed on one of the two Q500s. The detailed specifications are shown in Refs. [14, 25, 26], and the main parameters are summarized in Table 1. Fig. 2 shows the schematic layout of the superferric STQ. It has a 170 mm pole tip radius, which is relatively large compared to the magnetic pole length. The maximum field at the pole tip is 2.4 T.

The three-dimensional magnetic field maps of superferric STQs, STQ7, STQ11, STQ22, and STQ24, in the BigRIPS were measured over several years in intervals between beam times. Among these, STQ11 is the Q500-Q1000-Q500 type, and the others are Q500-Q800-Q500. Details of the measurement, including the field mapping device, are shown in Ref. [26]. The field mapping device is fixed to the up- and downstream flanges, and the horizontal and vertical positions are set to within 0.5 mm against the flange. The three-axis Hall sensors are placed on an arm that rotates in the $\theta$ direction, with $\theta = 0$ at the angle where the arm center aligns with the median plane of the STQ. Each three-axis Hall sensor is aligned so that the field in the $r$, $\theta$, and $z$ directions is each measured by one element of the sensor. Three sensors are mounted on the arm at radii of 81 mm, 94 mm, and 107 mm, and three-dimensional magnetic fields were measured at these radii simultaneously in 3 degree steps in the $\theta$ direction, and 10 mm steps in the $z$ direction. The Hall sensors used for the measurement were calibrated using NMR in a uniform field inside a dipole magnet before measurement. Field maps were measured at 165 A, the maximum current, and in steps of 10 A from 10 A to 160 A, and again by smaller current steps of 5 A or 3.3 A from 40 A to 120 A, where the effective length changes rapidly.

In Fig. 3, $\theta$-dependence of the measured magnetic fields of Q500 in STQ24 are shown as an example. The three components of the magnetic field, $B_r, B_\theta$ and $B_z$ at a radius of 107 mm with the excitation current of 100 A are plotted as functions of $\theta$ using blue, red, and green symbols, respectively. The left panel shows the result near the magnet center, and the right panel shows the result near the effective field boundary of the magnet. Four-fold symmetry is easily observed, and the quadrupole components are obviously the dominant components. $B_z$ is related to the $z$ derivative as shown in equation (3c), and thus it disappears near the magnet center but becomes large near the effective field boundary where the fringing field changes rapidly along the $z$ axis.

Firstly, let us consider multipole components of the magnetic field. The multipole components were extracted from the measured results at each point of $z$ with fixed $r$ by Fourier analysis with respect to $\theta$. We define the relative strength of the normal multipole components at $r$ = 107 mm normalized by the quadrupole strength at the center $z$ = 0,

$$R = \frac{B_{r,n}(z,\ r = 107\text{mm})}{B_{r,2}(z = 0,\ r = 107\text{mm})}.$$

We show, in Fig. 4, the relative strength $R$ at different positions of $z$ for $n$ = 1 to 10 except for $n$ = 2. Blue and red bars indicate the results at $z$ = +250 mm (near the effective field boundary) and z = 0 mm (magnet center), respectively. Among them, only the 12- and 20-pole components ($n$ = 6 and 10) are allowed in the exact four-fold symmetry. The strengths of the 12-pole component are around 0.4%, and the others are within 0.2% of the quadrupole components. A dipole component arises from the displacement between the rotating axis of the field mapping device and the symmetry axis of a magnet. The dipole components that appeared here corresponded to a displacement of about 0.3 mm at maximum. Sextupole and other higher multipoles are considered to be spurious, because they can be caused by measurement errors, for example, a deflection of the field mapping device. Since the strengths of these higher-multipole components are small and we expect that they do not affect ion-optics much, we focus only on the quadrupole component in the following analysis. To extract the normal quadrupole component, we shifted the origin of the azimuthal angle $\theta$. Since we have not used any reference magnet, a small misalignment of Hall sensor direction would generate artificially skewed quadrupole components. We redefined the origin of $\theta$ for each Hall sensor, such that skewed quadrupoles vanished. The maximum correction was 3 degrees.

The measured quadrupole field distributions for a Q500 magnet are shown in Fig. 5. The quadrupole components of the radial and azimuthal fields; $B_{r,2}$ and $B_{\theta,2}$ at a radius of 107 mm are shown as functions of $z$ with various excitation currents. The origin of the $z$-axis was redefined, so that the center of the magnet corresponds to $z$ = 0. The measured quadrupole distribution in the vicinity of the geometrical center was fitted by a quadratic function and we defined the magnetic center at the point where the value was maximum. When the excitation current is low, the curve is flat near the magnet center. As the current increases, the flat area disappears, and the fringing field is extended to the center. In order to see the difference between the radial and azimuthal components, the quadrupole field distributions $B_{r,2}$ and $B_{\theta,2}$ at $r$ = 107 mm at the excitation current of 100 A are plotted in Fig. 6 together with the $z$-component $B_{z,2}$. $B_{r,2}$ and $B_{\theta,2}$ mostly overlap but differ slightly near the effective field boundary. We can see from equations (3a) and (3b) that $B_{r,2}$ and $B_{\theta,2}$ would match if only the first order term ($m$ = 0) exists, so any difference arises from the pseudo-multipole term $b_{n,m}$ for $m \geq 1$, showing that these terms cannot be ignored. The distribution of the quadrupole components ($n$ = 2 and $m$ = 0) can be extracted correctly excluding these higher-order terms.

Using the formula in Sec. 2, the leading term of the quadrupole components is extracted. Recall the Fourier transform of the field map

$$\tilde{B}_{r(\theta),2}(k) = \int_{-\infty}^{\infty} B_{r(\theta),2}(z) e^{-ikz} dz.$$

We evaluate this transformation in a discrete form, since we performed the magnetic field measurement at finite points of $z$. Assuming that we measured the magnetic field at the discrete points $z_i$; for $i = 1, 2, \cdots, N$, with the interval of $\Delta z$, the above integral is replaced by the following discrete form.

$$\tilde{B}_{r(\theta),2}(k_j) = \sum_{i=1}^{N} B_{r(\theta),2}(z_i)\, e^{-ik_j z_i} \Delta z,$$

where $k_j$ is a discrete set defined as

$$k_j = \frac{2\pi}{N\Delta z}\, j \quad (\text{for } j = 0, 1, \cdots, N/2 - 1).$$

Rewriting equations (5) and (6) in discrete forms, a set of complex Fourier coefficients $\tilde{b}_{2,0}(k_j)$ is obtained by algebraic calculation. Performing the inverse Fourier transform in a discrete form, we obtain a discrete set of $b_{2,0}(z_i)$ from which we can evaluate the value of $b_{2,0}(z)$ at any point of $z$ by an appropriate interpolation or curve fitting. In Sec. 4, we discuss in detail the curve fitting using the Enge function for ion-optical calculations.

Once the leading term $b_{2,0}(z)$ is calculated, the higher order term $b_{2,m}(z)$ can be in principle obtained sequentially by the recurrence relation (2). Since the evaluation of repeated derivatives is numerically difficult, it is more practical to use the recurrence relation of the Fourier coefficients $\tilde{b}_{n,m}(k)$ given in Sec. 2.2. The values of the higher order terms $b_{2,m}(z)$ at the discrete $z_i$ can be calculated by the inverse Fourier transform of $\tilde{b}_{n,m}(k_j)$ as well as the leading term $b_{2,0}(z)$. In Fig. 7, $b_{2,0\cdots4}(z)$ obtained from the measurement result of $B_{\theta,2}$ at 100 A are plotted as an example, where the recurrence relation (2) was employed for the derivation. Here $b_{2,1}$ is relatively large, reaching 7%–8% of $b_{2,0}$ at maximum, whereas $b_{2,2}$, $b_{2,3}$, and $b_{2,4}$ are 1% −2%, 0.3%−0.5%, and 0.1% −0.2%, respectively. The terms with $m$ = 5 and larger were 0.1% or less, and thus the calculation of terms up to $m$ = 5 was sufficient to obtain 0.1% accuracy.

Fig. 8 shows the $b_{2,0}$ distribution extracted from the $B_{\theta,2}$ map measured at $r$ = 107 mm as an example. Here, the scale radius $r_0$ is assumed to be the warm bore radius of 120 mm. As the current increases, the flat areas at the magnet center decrease, and the shape of the distribution changes as well. Fig. 9 compares the distributions of $B_{r,2}$ and $B_{\theta,2}$ against the extracted $b_{2,0}$. The distributions of both $B_{r,2}$ and $B_{\theta,2}$ differ from the distribution of $b_{2,0}$. This difference indicates the effect of the pseudo-multipole terms $b_{2,m>0}$. For $B_{\theta,2}$, the reduction factor of $n/(n+2m)$ is applied, so the difference between $B_{\theta,2}$ and $b_{2,0}$ is smaller than that between $B_{r,2}$ and $b_{2,0}$. If the measurement conditions are the same for $B_{r,2}$ and $B_{\theta,2}$, the $b_{2,0}$ extracted from $B_{\theta,2}$ is expected to be more accurate.

In Fig. 10, the $b_{2,0}$ distributions extracted from data for $B_{r,2}$ and $B_{\theta,2}$ measured at three different radii at the same time are compared. They should reproduce the same three-dimensional magnetic field and match, and indeed they agree with each other within 0.1%. Furthermore, $B_{r,2}$ and $B_{\theta,2}$ were calculated from the extracted $b_{2,0}$ to confirm that the measurement results were reproduced. The results showed that they match on the order of 0.1%. These results demonstrate the validity of our procedure for extracting $b_{2,0}$.

## 4. Parameterization of field distributions

The ion-optical calculations, including the ion-optical search, are performed using the obtained $b_{2,0}(z)$ distributions. During the search calculations, we search for the excitation currents (or magnetic fields) that satisfy the required ion-optical conditions. Fig. 11 shows the field gradient and effective length as functions of the excitation current. Here, the effective length is defined as $L_{\text{eff}}=\int b_{2,0}(z)dz/b_{2,0}(z=0)$. Note that since the pseudo-multipole terms integrate to zero as mentioned above, these terms do not change the effective length of the quadrupole, so only the leading term is needed for the effective length. The excitation curve of the field gradient bends around 80 A, and the effective length changes significantly as well, revealing the effect of core saturation caused by a strong magnetic field. This change must be reflected correctly in the ion-optical calculation.

We parameterized the extracted $b_{n,0}(z)$ distribution using the Enge function [21], which has been widely used historically to describe the fringing field, and performed the ion-optical calculation. The code COSY INFINITY [17], which includes the fringing field in the form of the Enge function, was used for our ion-optical calculation. Such parameterization allows us to interpolate the fringing field distributions to the excitation currents for which magnetic fields are not measured, facilitating the ion-optical calculations. From the discussion in Sec. 2, the three-dimensional magnetic field which satisfies Maxwell's equation in the region under consideration can be completely determined by $b_{n,0}(z)$. Since it is only a function of the z position, it is much easier to parameterize the $b_{n,0}(z)$ function than to parameterize the full three-dimensional field directly. It is also of great advantage in extracting the $b_{n,0}(z)$.

The Enge function is defined as follows, normalized by the magnetic field strength at the magnet center.

$$B(\zeta)/B_c = F(\zeta) \equiv \frac{1}{1+\exp[\sum_{k=1}^{N} a_k (\zeta/D)^{k-1}]}.$$

Here, $\zeta$ denotes the position along the beam axis from the effective field boundary, and $D$ denotes the full aperture of the magnet. $B_c$ indicates the magnetic field strength at the magnet center. The parameter number $N$ must be an even number, and the highest-order coefficient $a_N$ must be positive so that $F(\zeta)$ asymptotically reaches 1 as $\zeta$ goes to $-\infty$ and 0 as $\zeta$ goes to $+\infty$. In this case, the Enge function represents the magnet's exit side, or the fringing field $F_{out}$ from the inside to the outside of the magnet. By using $F_{in} = F(-\zeta)$ at the magnet's entry side, the Enge function that results in 0 at $\zeta = -\infty$ and 1 at $\zeta = +\infty$ can be obtained. For magnets that are symmetric at the entry and exit sides, the coefficients $a_k$ become the same for $F_{in}$ and $F_{out}$. We used the $N$ = 6 Enge function for which COSY is equipped. In many cases, it provides sufficient accuracy.

For the ion-optical calculation for BigRIPS, the distribution of the extracted $b_{2,0}(z)$ is fitted using the Enge function for both the entry and exit sides, and the coefficients $a_{1,\cdots,6}$ are deduced. When a magnet is sufficiently long compared to the aperture and the distribution at the center is flat, simply connecting the Enge function of the entry and exit sides at the magnet center does not become a serious problem. However, for the large-aperture magnet we are discussing here, the fringe region reaches the magnet center, and the Enge function changes slightly even at $z$ = 0. Therefore, if we connect the two Enge functions at $z$ = 0, the $z$ derivative becomes discontinuous at $z$ = 0. In such a case, the pseudo terms $b_{2,m\geq 1}(z)$ diverge at $z$ = 0 as can be seen from the recurrence relation (2), and thus the resulting 3D fields diverge at $z$ = 0. In order to avoid this problem, we used the following folded $b_{2,0}(z)$ [27],

$$b_{2,0}(z) = B_c \cdot F_{in}\big(-(z + L/2)\big) \cdot F_{out}\big(+(z - L/2)\big), \quad (7)$$

where $L$ is a nominal length of the magnet. We modified COSY to use this form of the function.

Fitting to the Enge function was performed for each excitation current. Fig. 12 illustrates the Enge coefficients as functions of excitation currents, which were obtained for Q500 quadrupole magnets. The upper six are the Enge coefficients $a_{1,\cdots,6}$ of the entry side, and the lower six are those of the exit side. We obtained $b_{2,0}$ at an arbitrary excitation current value by interpolating these Enge coefficients. For the interpolation, we sometimes fit the Enge coefficients using polynomials.

In order to study the sensitivity of the Enge function, we examined the change in the Enge function when each Enge coefficient was increased by 10% from an optimal value. The most sensitive coefficient was $a_2$, and the Enge function changed by 2% at maximum for a 10% change. Other coefficients changed by 0.1% or less. This is because the $a_2$ component is dominant in the $|\zeta/D| < 1$ domain, which determines the shape of the fringing field.

It is important to determine the most sensitive coefficient $a_2$ accurately. The position of a magnet along the beam axis is one of the most important points to consider. For our magnets, when the position changes by 1 mm along the beam axis near the effective field boundary, the magnetic field strength changes by 1%. When fitting with the Enge function, this change appears as a fluctuation of coefficient $a_2$. Because the magnetic field was measured at 10 mm intervals, measurement was not always symmetric with respect to the magnet center. In order to minimize the fluctuation of $a_2$, we fitted the extracted $b_{2,0}(z_i)$ distribution in the vicinity of the geometrical center with a quadratic function and defined the magnetic center as the point where the value was at maximum. In addition, for Q800 and Q1000, which are in the middle of the triplet magnets, we assume that the magnetic field distributions are symmetric about the center and use the same Enge coefficients at the entry and exit sides.

In addition to the superferric-type STQs for which the magnetic fields were measured, the air-core type STQ1 is used in the BigRIPS. Because there are no magnetic field measurement data for this magnet, we performed a similar magnetic field analysis based on the TOSCA calculations and extracted the $b_{2,0}$ distributions. We found that from the effective field boundary to the outside, a large undershoot to the negative side occurred, and from the effective field boundary to the center, there were areas where the magnetic field was stronger than that at the center. Because the Enge function takes only values between 0 and 1, this type of field distribution cannot be described. Therefore, we need to add a correction term to the Enge function and used the following functional form.

$$F(\zeta) = \frac{1}{1+\exp(a_1 + a_2\cdot(\zeta/D) + \cdots + a_6\cdot(\zeta/D)^5)} + a_7\tanh\big(a_8 + a_9\cdot(\zeta/D)\big)\cdot\exp\left[-\left(\frac{\zeta/D + a_{10}}{a_{11}}\right)\right].$$

## 5. Ion-optical calculation of the BigRIPS

In the ion-optical search calculations, we divide the BigRIPS ion-optical system into subsystems separated by foci. Details of the configuration of the BigRIPS can be found in Refs. [11-13]. In each optical subsystem, we require a certain ion-optical condition expressed as the values of transfer matrix elements summarized in Table 2. These conditions are required by the ion-optical specifications of the BigRIPS as the spectrometer, such as momentum resolutions, angular- and momentum-acceptances, the maximum magnetic rigidity $B\rho$ and so on.

In a standard ion-optics calculation using transfer matrices, the ion-optical conditions determine the “normalized field gradient (field gradient/magnetic rigidity) [2]” of the quadrupole magnets with fixed effective lengths. For a given magnetic rigidity of the reference particle, the field gradient of each magnet is determined by the normalized field gradient solved by an ion-optical code. The excitation current is then determined by the excitation curve of each quadrupole. In contrast the present ion-optical calculations with the STQs are not simple. The effective length and the field distribution with large fringing fields drastically change according to the excitation current. The excitation curves are far from a straight line due to the strong saturation of the iron. Under these circumstances an elaborate method is particularly needed for the ion-optical search calculations.

We adopt the following procedure for this purpose. For a given magnetic rigidity $B\rho$ of the reference particle, we search the field strengths of the quadrupoles at the center so as to satisfy the required ion-optical conditions, which are given in Table 2 as transfer matrices, by using the code COSY INFINITY with the field distributions $b_{2,0}(z)$ discussed in Sec. 4. Repeating the COSY calculations by varying the rigidity of the reference particle $B\rho_{\mathrm{ref}}$, we obtain curves of the field strength for each quadrupole as functions of $B\rho_{\mathrm{ref}}$. Then, together with the excitation curve shown in Fig. 11, we determine the excitation current setting for the rigidity of the beam in which we are interested.

Now we discuss our ion-optics calculation using the code COSY INFINITY in more detail. For a given value of the field strength at the center of the quadrupole magnet, the field distribution $b_{2,0}(z)$ is uniquely determined by the parameterization (7) in Sec. 4. The code COSY INFINITY calculates the transfer map of the quarupole magnet using the field distribution $b_{2,0}(z)$ thus obtained, in which the effects of the fringing field are properly taken into account. We calculate the transfer maps for all the measured field distributions at various excitation currents and store them in files in advance. With these stored transfer maps, we can then search for the field strength of each quadrupole magnet to satisfy the ion-optical requirement listed in Table 2. This method is called "fringe mode 2 (FR2)" [17, 28]. The search is performed using the fit procedure in the code COSY INFINITY.

Fig. 13 shows search results of field strengths for all the BigRIPS quadrupoles STQ1~STQ14 over the required ion-optical conditions. Blue, purple and green lines represent upper-, middle- and downstream quadrupoles in the STQ, respectively, and the field strengths at the warm bore radius of 120 mm are plotted as functions of the rigidity $B\rho$. For all quadrupoles, the field strengths are fitted by linear functions of the $B\rho$ value. The slope and offset parameters are stored in a configuration file on the magnet control system. Together with the relation between the field gradient and the excitation current, which is shown in Fig. 11, all the excitation currents can be determined by just specifying the $B\rho$ value of the reference particle in the central orbit. Other configuration files are prepared for different ion-optical modes, if necessary. For a specific ion-optical mode, the magnets are set by loading the file of the corresponding mode onto the control system.

The final calculation of the transfer matrix elements was performed at the specific $B\rho$ value for the experiments by using the Enge coefficients fitted with a polynomial function of the excitation current or the field gradient, obtained in Sec. 4. Using the COSY's DA (differential algebra) based numerical integrator [17, 28], the transfer map is computed to the full accuracy of the integrator with a realistic field profile given by the folded function (7). To examine the fundamental optical properties of the BigRIPS and the predictability of COSY calculation, we compare the calculated transfer matrix elements with those extracted from the actual beam trajectories measured during the BigRIPS beam time.

In the BigRIPS separator, the positions and angles of the particles were measured for each event using position sensitive detectors called PPACs [29, 30] installed in every focal plane chamber. The details of the experimental setup can be found in Ref. [15]. The initial and final phase space variables are related to the matrix elements by a linear transformation in the first-order optics:

$$\begin{cases} x_f = (x|x)x_i + (x|a)a_i + (x|\delta)\delta, \\ a_f = (a|x)x_i + (a|a)a_i + (a|\delta)\delta, \end{cases}$$

where $x$, $a$ and $\delta$ are the horizontal position, horizontal angle, and deviation of the magnetic rigidity of the particle with respect to the central ray. Suffixes $i$ and $f$ indicate initial and final points, respectively. We extract the $(x|x)$ element, for example, as follows. We plot measured initial and final beam positions in the $(x_i, x_f)$ plane with narrow gates of $a_i$ ~ 0 and $\delta$ ~ 0, and the slope of the linear fit corresponds to the $(x|x)$ element. Other transfer matrix elements were obtained similarly. The extracted transfer matrix elements of F3-F5, F5-F7 and F3-F7 measured by secondary beams are compared with the COSY calculation in Fig. 14. Here F3, F5 and F7 indicate the foci in the BigRIPS separator (see Refs. [11-13,15] for the details). The red and blue bars indicate the measured ones and the COSY calculation, respectively. The green bars indicate the values used as the fitting conditions in the field strength search. Diagonal elements $(x|x), (a|a)$ and dispersion terms $(x|\delta), (a|\delta)$ are well reproduced by the COSY calculation, while disagreements of off-diagonal elements $(x|a)$ and $(a|x)$ are relatively large. They are essentially small values under the standard ion-optical mode since the point-to-point and parallel-to-parallel transport system is designed. The matrix element $(x|a)$ is very sensitive to the strengths of the quadrupole magnets. In other words, the focusing condition of the ion-optical system is sensitive to the excitation currents of the quadrupoles. Therefore even small errors in magnet settings result in poor focus of the ion-optical system. The focusing is very essential for the effectiveness of slits and wedge-shaped degraders for the isotope separator. In particular, in the procedure of trajectory reconstruction for improving particle identification powers [15], measurement errors in angles can be minimized when the focusing condition $(x|a) = 0$ is attained. In actual operations of the BigRIPS we always check the focusing at each focus and adjust the excitation currents of the quadrupole magnet to obtain good focus during the secondary beam tuning of the experiment. The focus tuning is carefully performed in the achromatic sections not to disturb the dispersions. The amount of tuning is only a few percent in most cases. We aim at precise ion-optical settings without any tuning.

## 6. Conclusion

We performed a three-dimensional magnetic field analysis of the superconducting quadrupole magnets used in the BigRIPS at RIBF as an example of large-aperture short-length quadrupole magnets with a large fringing field region, which varies drastically according to excitation current.

We extracted the first-order quadrupole field distributions from the field-map measurements of the surface of a cylinder and used them to perform a first-order ion-optical calculation of the BigRIPS.

For the magnetic field analysis, we developed a novel practical numerical method of extracting the leading term $b_{2,0}$ of the quadrupole components from the cylindrical surface measurements and reconstructing the full three-dimensional quadrupole fields from the $b_{2,0}$. We extracted $b_{2,0}$ from the field-map data measured at different radii and also from both $B_{r,2}$ and $B_{\theta,2}$ components. The obtained $b_{2,0}$ are independent of the radius and agree with each other within 0.1%, demonstrating the validity of our method. This method can be applied to multipole fields as well.

The obtained magnetic field distributions were parameterized using the Enge function. This allowed us to interpolate the field distributions to the excitation currents for which the field measurement was not performed and to calculate the first-order ion-optics for any current values using the code COSY INFINITY.

The method we used has various advantages; it simplifies the analysis because the field distributions at all positions can be obtained by extracting $b_{2,0}(z)$ from the measurement of the field distribution along the beam axis at one radius only. Moreover, the extracted field distribution consists of a $z$ function only and is easily parameterized. It is also convenient because the ion-optical calculation can be performed at all currents by interpolation.

We used the results obtained by this method for the optical design of the BigRIPS and the current setting of the STQs. The set current values required only a few percent of adjustment and are useful for quickly providing RI beams, including high-precision particle identification.

**Acknowledgments**

The authors would like to thank Dr. Y. Yano, RIKEN Nishina Center, for his support and encouragement. J.N. was supported by the U.S. Department of Energy, Office of Nuclear Physics, under Contract No. DE-AC02-06CH11357. T.K. is grateful to Dr. J. Stasko for his careful reading of the manuscript.

Table 1. Specifications of superferric quadrupole magnets which consist of STQs in the BigRIPS separator. They are characterized by large pole tip radii and high pole-tip fields (2.4 T).

| | Q500 | Q800 | Q1000 |
|---|---|---|---|
| Effective length [m] | 0.54 | 0.84 | 1.04 |
| Pole tip radius [mm] | 170 | 170 | 170 |
| Warm bore radius [mm] | 120 | 120 | 120 |
| Maximum pole-tip field [T] | 2.4 | 2.4 | 2.4 |
| Maximum field gradient [T/m] | 14.1 | 14.1 | 14.1 |

Table 2. Required conditions for the standard ion-optical mode in the BigRIPS separator. Here Fn denotes the foci in the BigRIPS separator (see Refs. [11-13,15] for the details). Note that the F5-F6-F7 subsystem is the mirror-symmetric setting of the F3-F4-F5 subsystem. The notation for the transfer matrix elements is based on the code COSY INFINITY [17], where $x$, $a$, and $\delta$ denote the horizontal position, horizontal angle, and fractional momentum deviation, respectively, and $y$ and $b$ the vertical position and angle, respectively. The dimensions in this table are given in mm for $x$ and $y$, mrad for $a$ and $b$, and percent for $\delta$.

| subsystem | F0-F1 | F0-F2 | F2-F3 | F3-F4 | F3-F5 |
|---|---|---|---|---|---|
| | $(x\|a) = 0$ | $(x\|a) = 0$ | $(x\|a) = 0$ | $(x\|a) = -0.791$ | $(x\|a) = 0$ |
| | $(y\|b) = 0$ | $(y\|b) = 0$ | $(y\|b) = 0$ | $(y\|b) = -0.272$ | $(y\|b) = 0$ |
| | $(x\|\delta) = -21.4$ | $(x\|\delta) = 0$ | $(a\|x) = 0$ | $(x\|\delta) = -22.1$ | $(x\|\delta) = 31.7$ |
| | $(a\|\delta) = 0$ | $(a\|\delta) = 0$ | $(b\|y) = 0$ | $(a\|\delta) = 0$ | $(a\|\delta) = 0$ |
| | $(x\|x) = -1.7$ | $(x\|x) = 2.0$ | $(x\|x) = -1.08$ | $(x\|x) = -1.4$ | $(x\|x) = 0.92$ |
| | $(y\|y) = -5.0$ | $(y\|y) = 1.6$ | $(y\|y) = -1.18$ | $(y\|y) = -3.45$ | $(y\|y) = 1.06$ |
| Searched STQs | STQ1, 2 | STQ3, 4 | STQ5, 6 | STQ7, 8 | STQ9, 10 |
| note | | after searching F0-F1 | | | after searching F3-F4 |

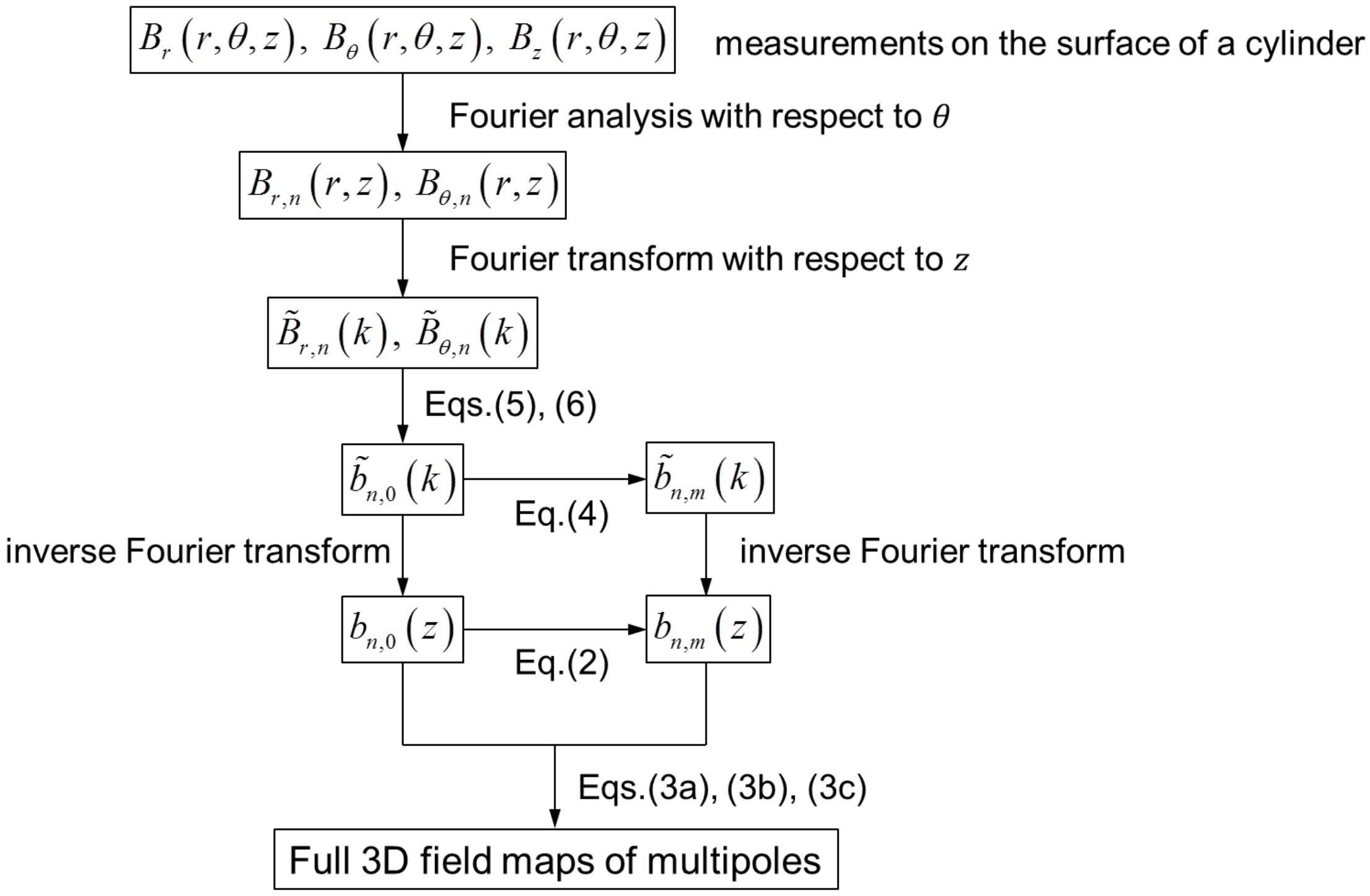

Fig. 1. Extraction of full 3D field maps of magnetic multipoles from 2D surface measurements. The process of extracting the leading term $b_{n,0}(z)$ and the pseudo terms $b_{n,m}(z)$ from the 2D measurements of the surface of a cylinder are shown step-by-step. The equations shown by the arrows indicate those used for the corresponding processes in Section 2.

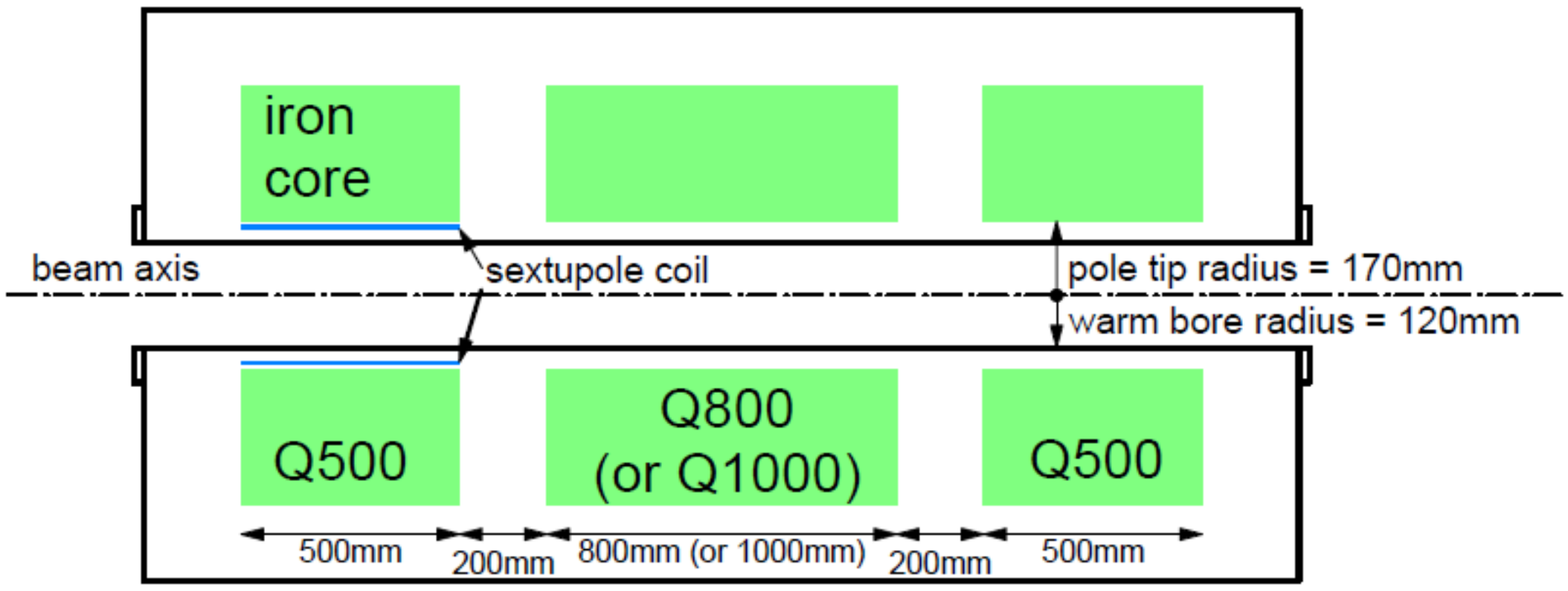

Fig. 2. A schematic layout of the superferric STQ in the BigRIPS separator, which consists of three large-aperture superconducting quadrupoles.

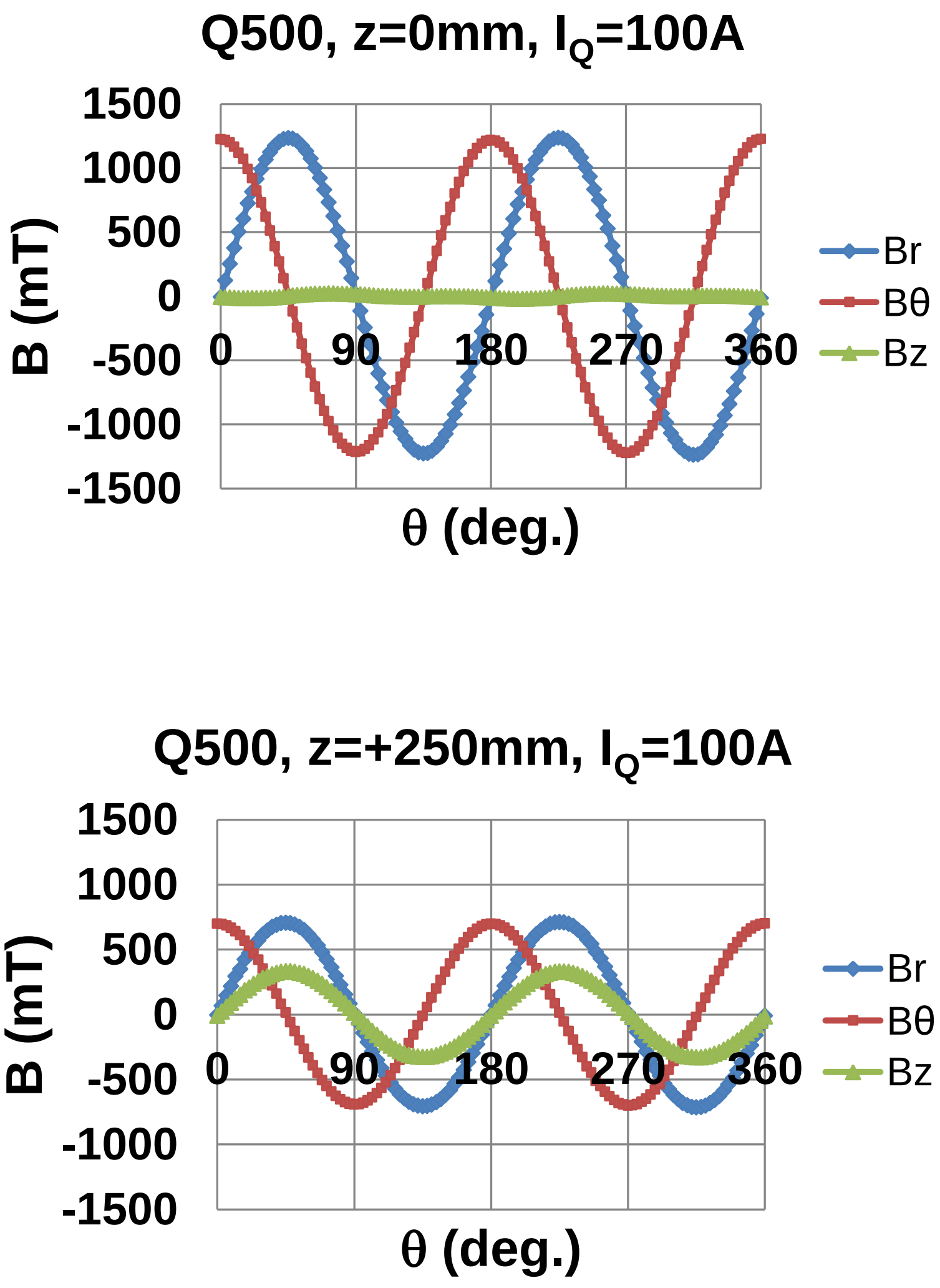

Fig. 3. Examples of field map measurements of a Q500 quadrupole magnet in STQ24. The radial-, azimuthal-, and *z*-components measured at a radius of 107 mm and at an excitation current of 100 A are plotted in blue, red, and green symbols as a function of azimuthal angle, respectively. Upper panel shows the results at $z$ = 0 mm (the center of the magnet); lower panel shows the results at $z$ = +250 mm (near the effective field boundary).

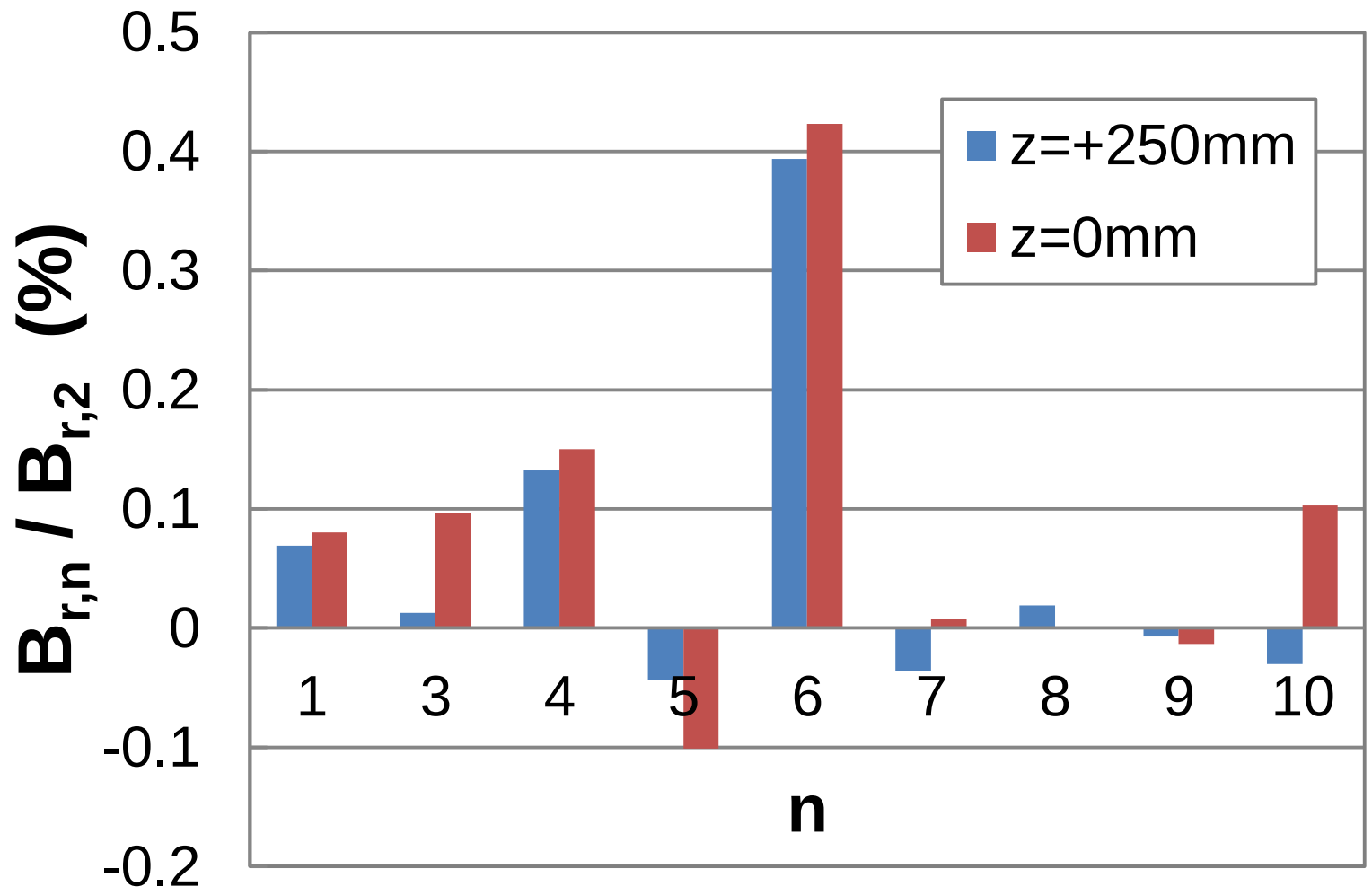

Fig. 4. Relative strengths of the multipole components normalized to the quadrupole component at the magnet center, expressed as $\boldsymbol{R} = \boldsymbol{B}_{\boldsymbol{r},\boldsymbol{n}}(\boldsymbol{z}, \boldsymbol{r} = \mathbf{107}\text{mm})/\boldsymbol{B}_{\boldsymbol{r},\mathbf{2}}(\boldsymbol{z} = \mathbf{0}\text{mm}, \boldsymbol{r} = \mathbf{107}\text{mm})$, which were obtained for a Q500 quadrupole magnet in STQ24. Blue and red bars indicate the ratios at $z$ = +250 mm (near the effective field boundary) and $z$ = 0 mm (at the center of the magnet), respectively.

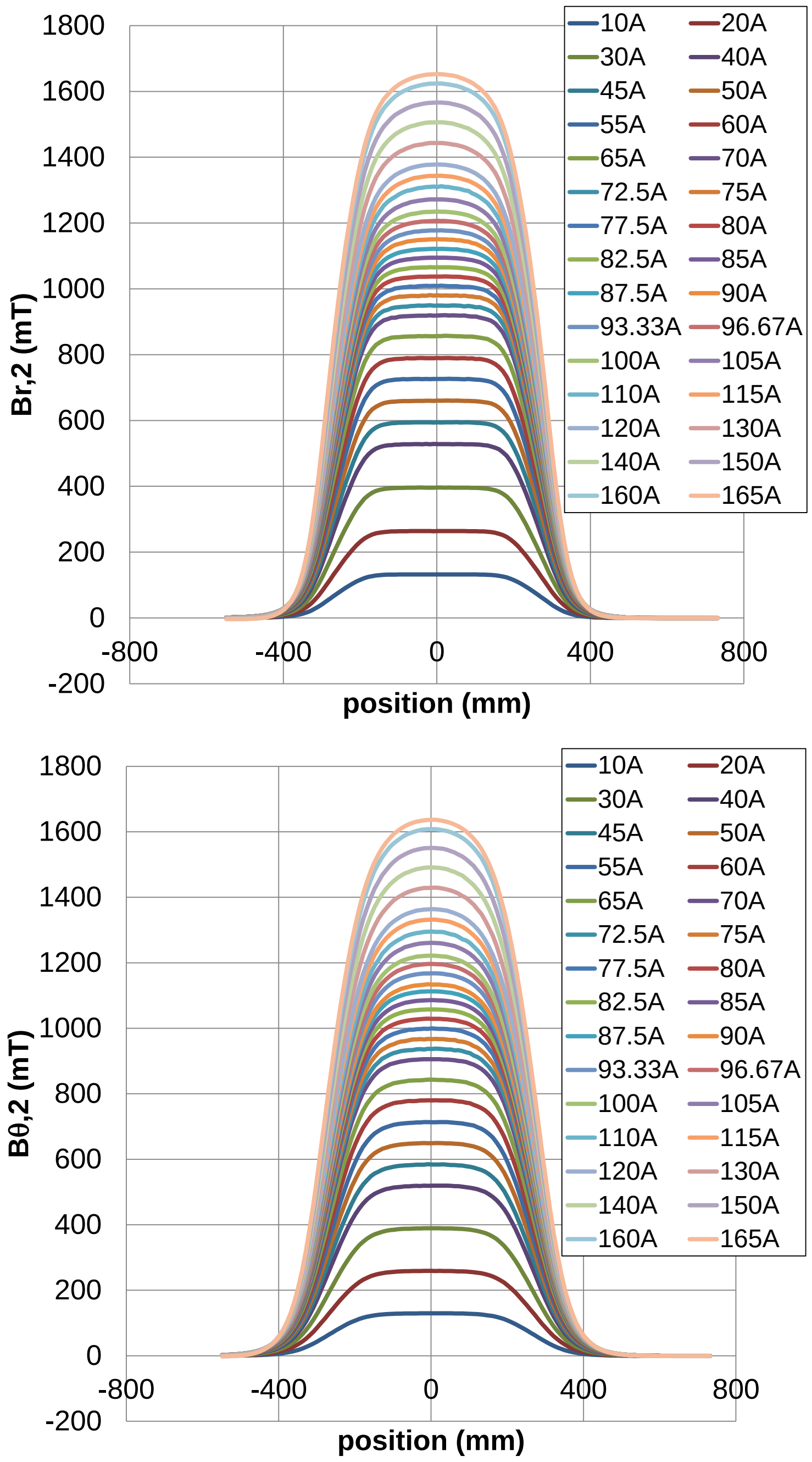

Fig. 5. Quadrupole field distributions extracted from the magnetic field map data, which were measured at a radius of 107 mm with various excitation currents for a Q500 quadrupole magnet in STQ24. Upper panel shows the radial component $\boldsymbol{B_{r,2}}$; lower panel shows the azimuthal component $\boldsymbol{B_{\theta,2}}$ as a function of the position *z* along the beam axis.

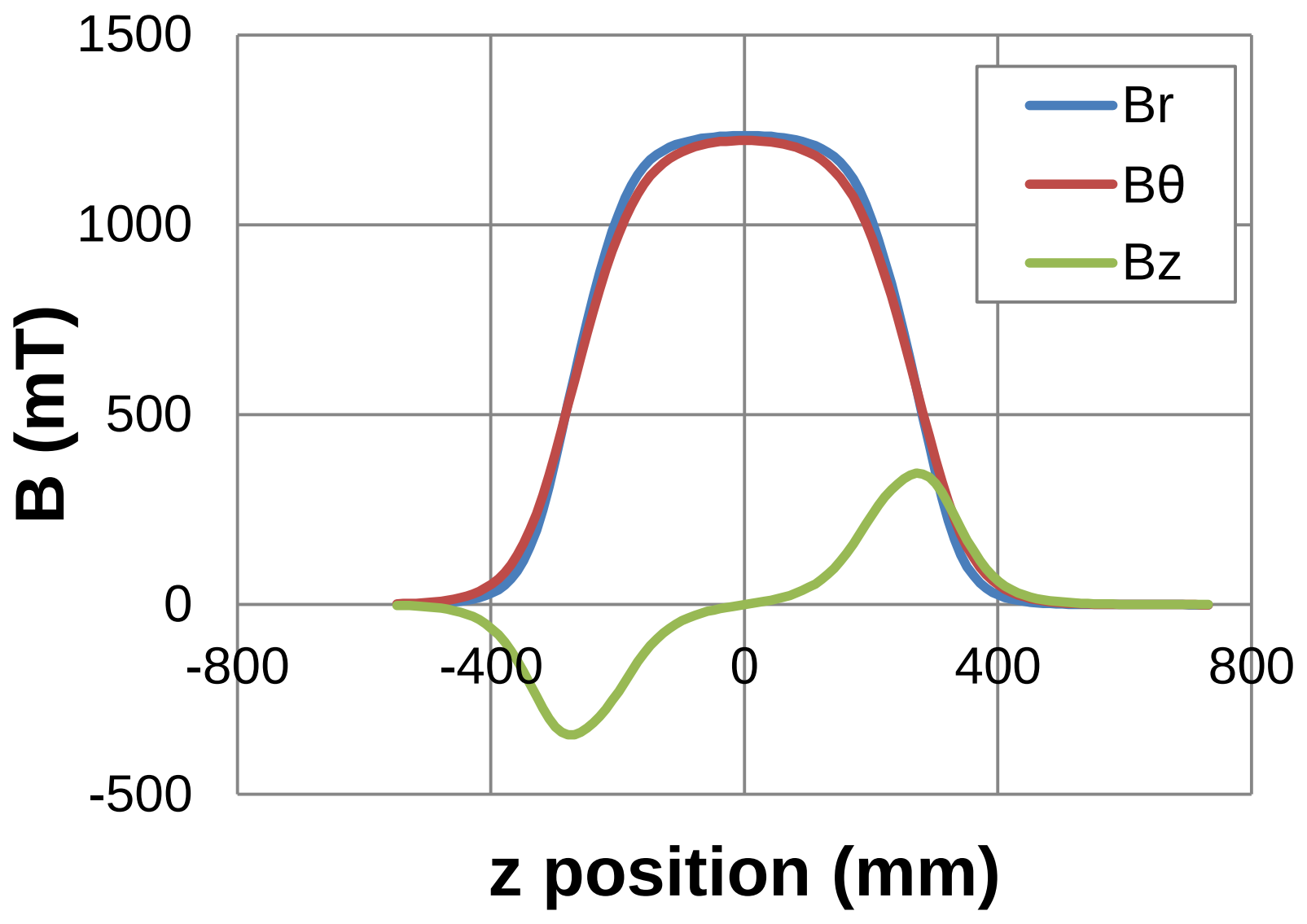

Fig. 6. Radial-, azimuthal-, and *z*-components of the quadrupole field $\boldsymbol{B_{r,2}}, \boldsymbol{B_{\theta,2}}$, and $\boldsymbol{B_{z,2}}$, which were measured at a radius of 107 mm for a Q500 quadrupole magnet in STQ24. $\boldsymbol{B_{r,2}}$ **and** $\boldsymbol{B_{\theta,2}}$ mostly overlap but differ slightly near the effective field boundary. They should match if only the leading term of $m$ = 0 exists, and so the figures indicate the effect of the pseudo terms of $m \geq 1$ in the measurement results.

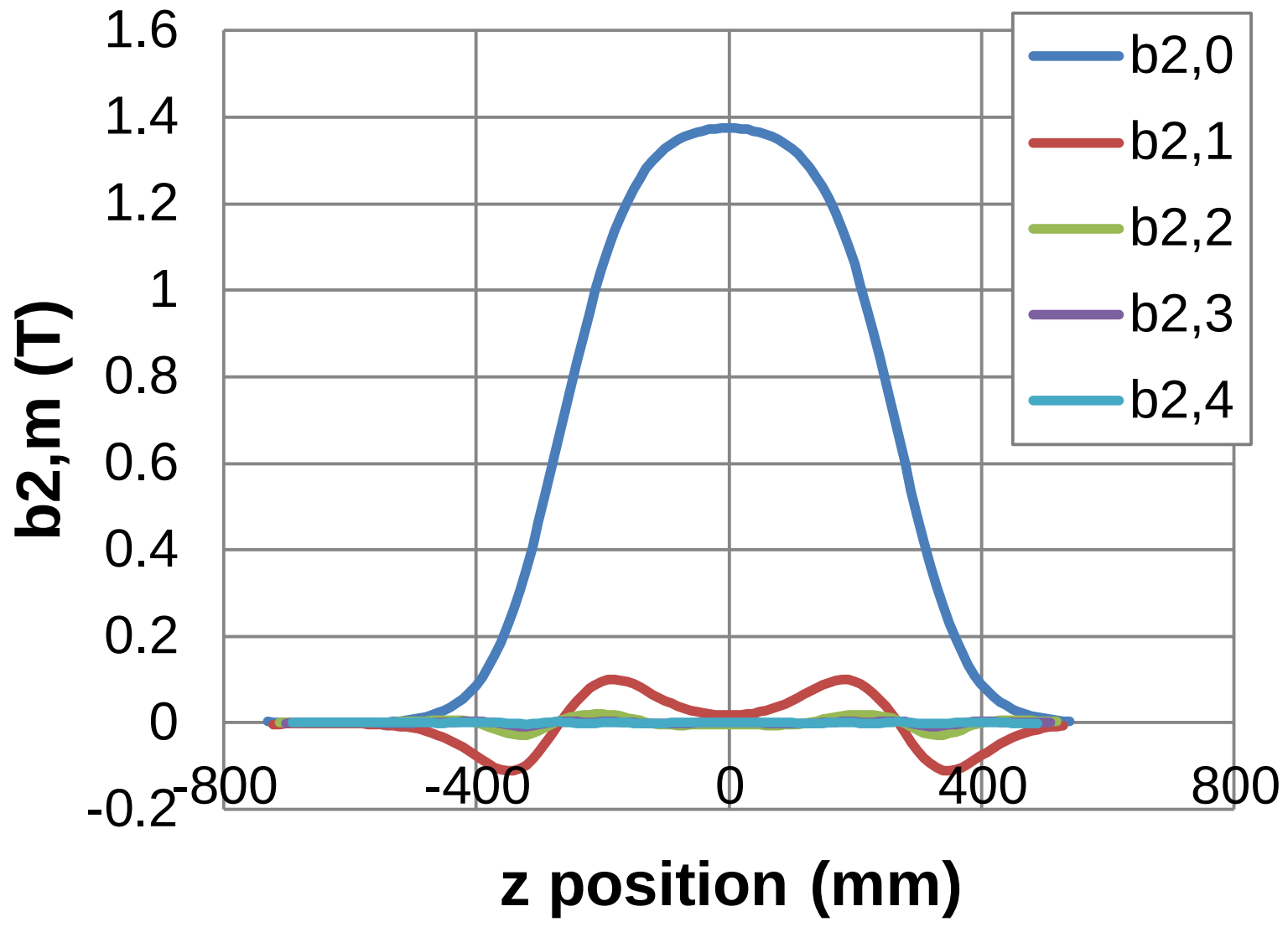

Fig. 7. Examples of $\boldsymbol{b}_{2,0\cdots4}(\boldsymbol{z})$ distributions. Here $\boldsymbol{b}_{2,0}$ was extracted from $\boldsymbol{B}_{r,2}$, which was measured at a radius of 107 mm and at an excitation current of 100 A for a Q500 quadrupole magnet in STQ24. The scale radius $\boldsymbol{r}_0$ is taken to be 120 mm (warm bore radius of STQ) to deduce $\boldsymbol{b}_{2,0}$. Pseudo terms $\boldsymbol{b}_{2,1\cdots4}$ were calculated from $\boldsymbol{b}_{2,0}$ with the recursion relation (2). $\boldsymbol{b}_{2,1}$ is relatively large, reaching 7%–8% of $\boldsymbol{b}_{2,0}$ at maximum. $\boldsymbol{b}_{2,2}, \boldsymbol{b}_{2,3}$, and $\boldsymbol{b}_{2,4}$ are 1%–2%, 0.3%–0.5%, and 0.1%–0.2%, respectively. The terms with $m$ = 5 or larger were 0.1% or less. See text.

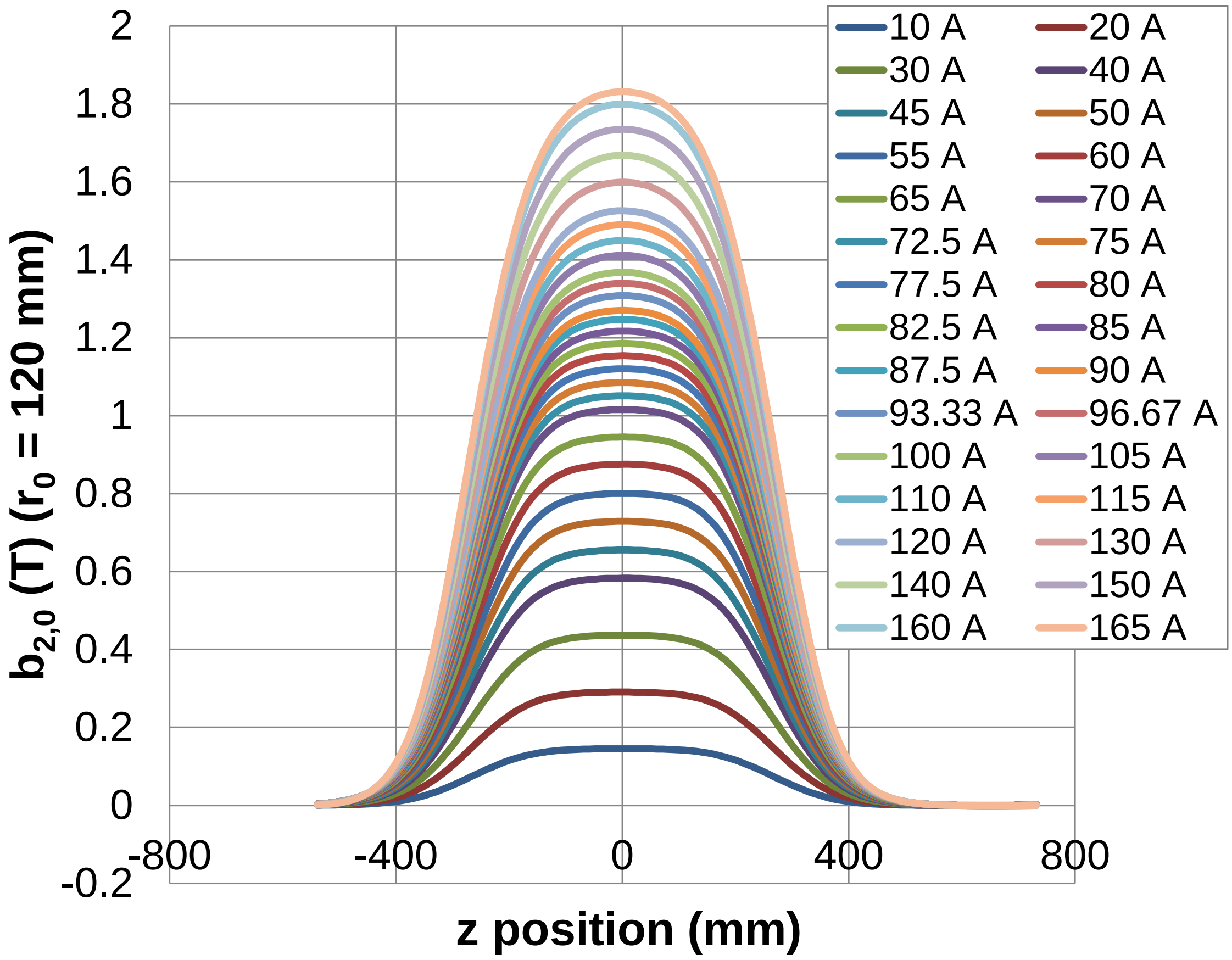

Fig. 8. $\boldsymbol{b_{2,0}(z)}$ at different excitation currents. Data are extracted from $\boldsymbol{B_{\theta,2}}$ distribution which was measured at a radius of 107 mm for a Q500 quadrupole magnet in STQ24. The scale radius $\boldsymbol{r_0}$ is taken to be 120 mm (warm bore radius of STQ). As the current increases, the flat areas at the magnet center decrease, and the shape of the distribution changes.

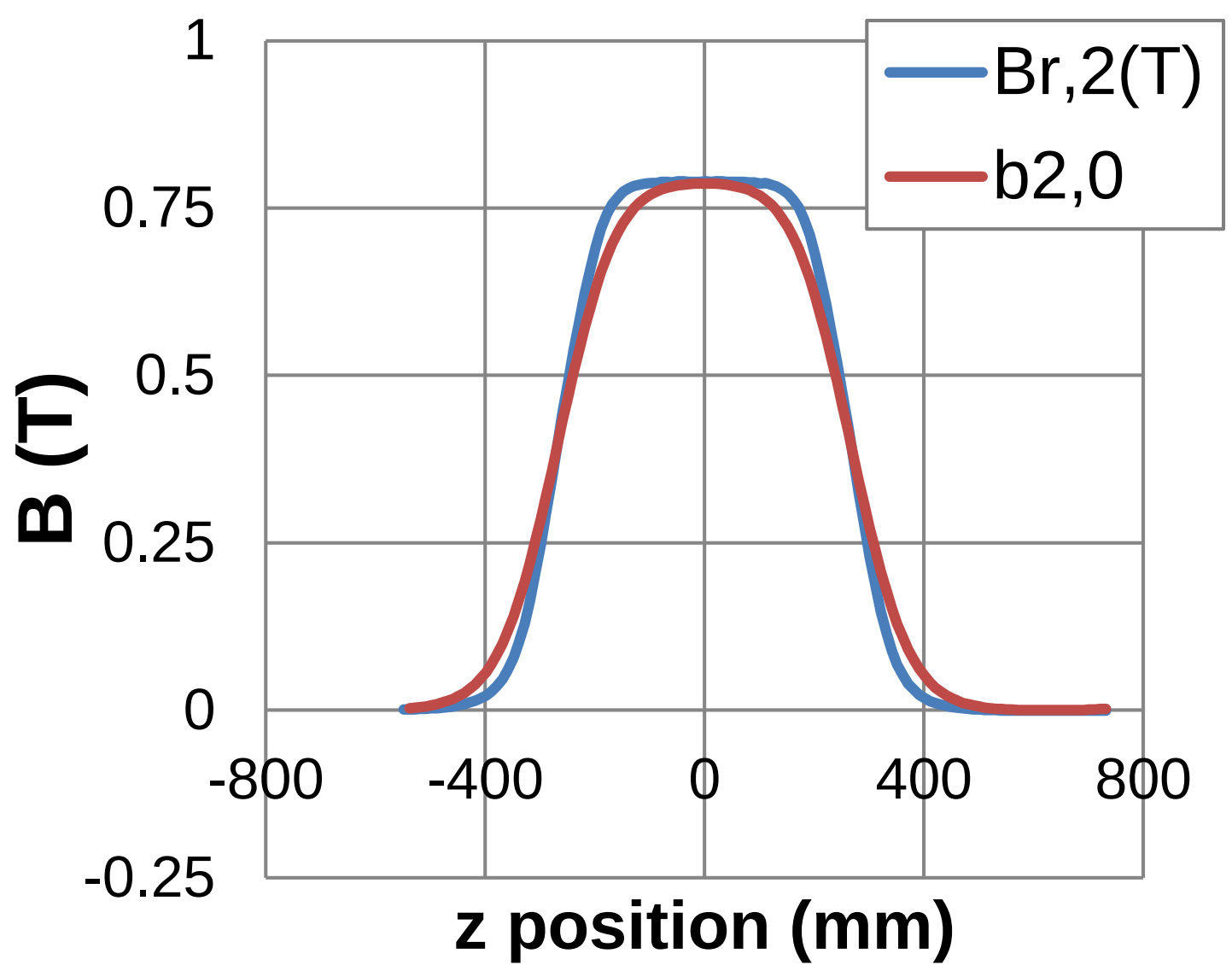

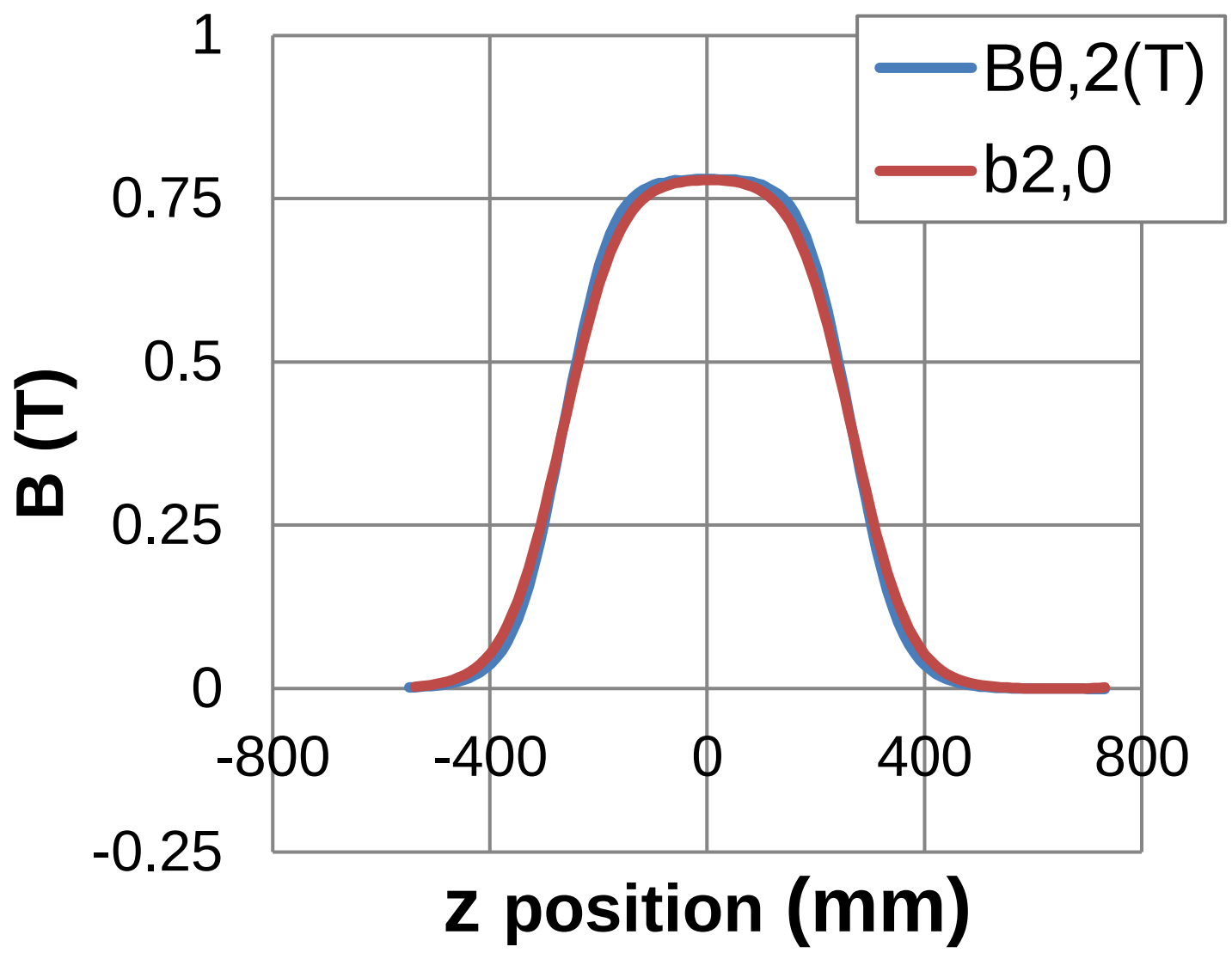

Fig. 9. Comparison of extracted $\boldsymbol{b_{2,0}}$ distribution with those of $\boldsymbol{B_{r,2}}$(upper) and $\boldsymbol{B_{\theta,2}}$(lower) which were measured at a radius of 107 mm for a Q500 quadrupole magnet in STQ24. Here the scale radius $\boldsymbol{r_0}$ is taken to be 107 mm (measurement radius), so that the $\boldsymbol{b_{2,0}}$ and $\boldsymbol{B_{r,2}}, \boldsymbol{B_{\theta,2}}$ can become the same values at the center ($z$ = 0). The distributions of both $\boldsymbol{B_{r,2}}$ and $\boldsymbol{B_{\theta,2}}$ differ from that of $\boldsymbol{b_{2,0}}$, indicating the effect of the pseudo terms $\boldsymbol{b_{2,m\geq 1}}$. For $\boldsymbol{B_{\theta,2}}$, the reduction factor $\boldsymbol{n/(n+2m)}$ is applied to the pseudo terms, and so the difference from $\boldsymbol{b_{2,0}}$ is smaller than that of $\boldsymbol{B_{r,2}}$.

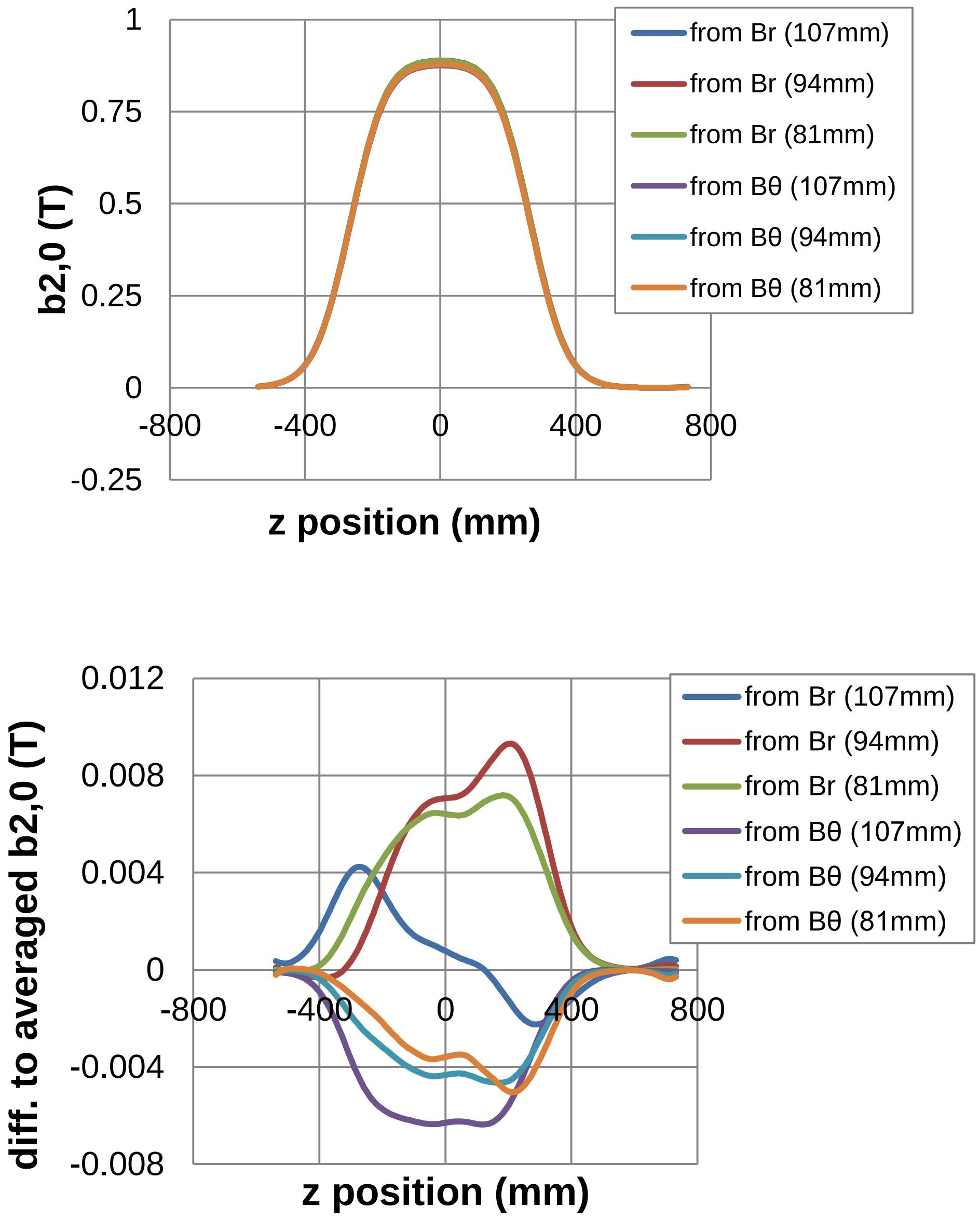

Fig. 10. $\boldsymbol{b_{2,0}}$ extracted independently from $\boldsymbol{B_{r,2}}$ and $\boldsymbol{B_{\theta,2}}$, which were obtained at three different radii for a Q500 quadrupole magnet in STQ24. Here the scale radius $\boldsymbol{r_0}$ is commonly taken to be 120 mm (warm bore radius of STQ). Upper panel shows the six $\boldsymbol{b_{2,0}}$ distributions thus obtained, which overlap one another. Lower panel shows the difference between the averaged $\boldsymbol{b_{2,0}}$ and each $\boldsymbol{b_{2,0}}$ for ease of viewing. Since they should reproduce the same three-dimensional magnetic field, all the $\boldsymbol{b_{2,0}}$ should ideally be the same. Indeed, the figure shows that they agree within 0.1% in practice.

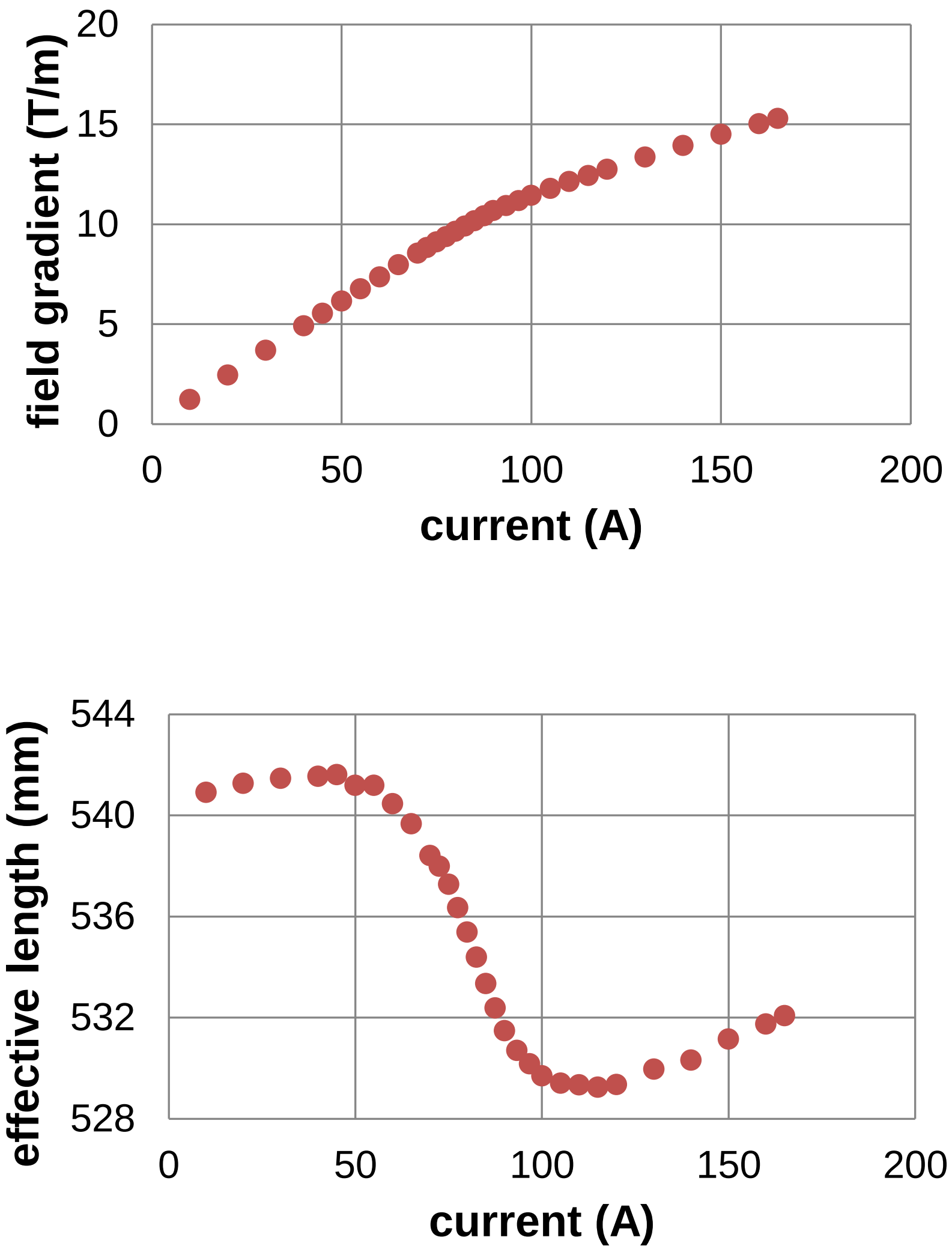

Fig. 11. Field gradient (upper) and effective length (lower) plotted as a function of excitation current, which were measured for a Q500 quadrupole magnet in STQ24. The field gradient starts to curve at around 80 A, and the effective length changes significantly, revealing the effect of core saturation caused by a strong magnetic field.

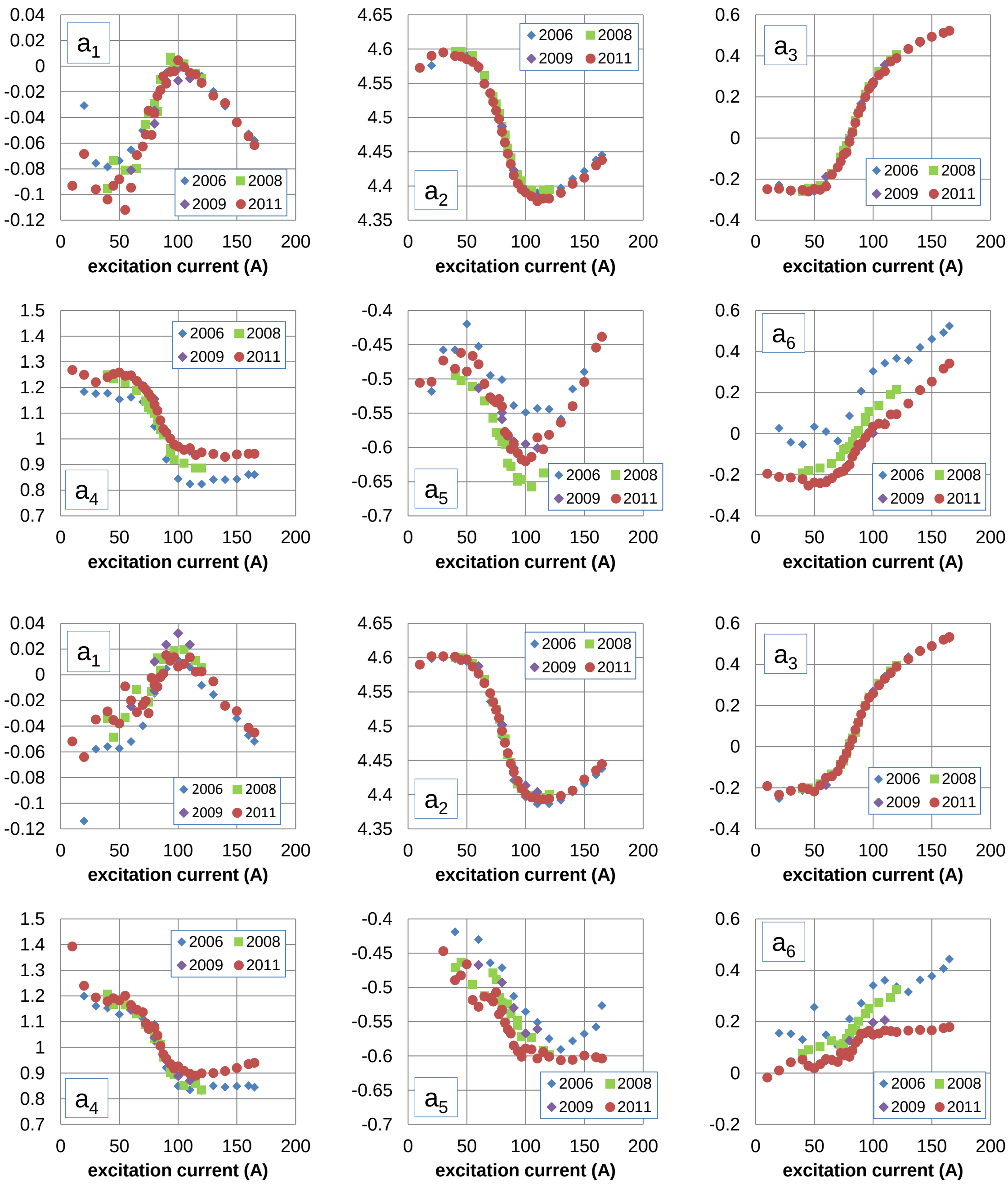

Fig. 12. Enge coefficients $\boldsymbol{a_1}, \cdots, \boldsymbol{a_6}$ plotted as functions of excitation currents, which were obtained for Q500 quadrupole magnets by fitting the discrete set of $\boldsymbol{b_{2,0}(z_i)}$ with the Enge function. Upper six are $\boldsymbol{a_1}, \cdots, \boldsymbol{a_6}$ at the entry side; lower six are those at the exit side. Four groups of symbols indicate the year when we measured the field-map data: The measurement was perfromed for STQ7 and STQ11 in 2006, STQ22 in 2008 and 2009, and STQ24 in 2011. See text.

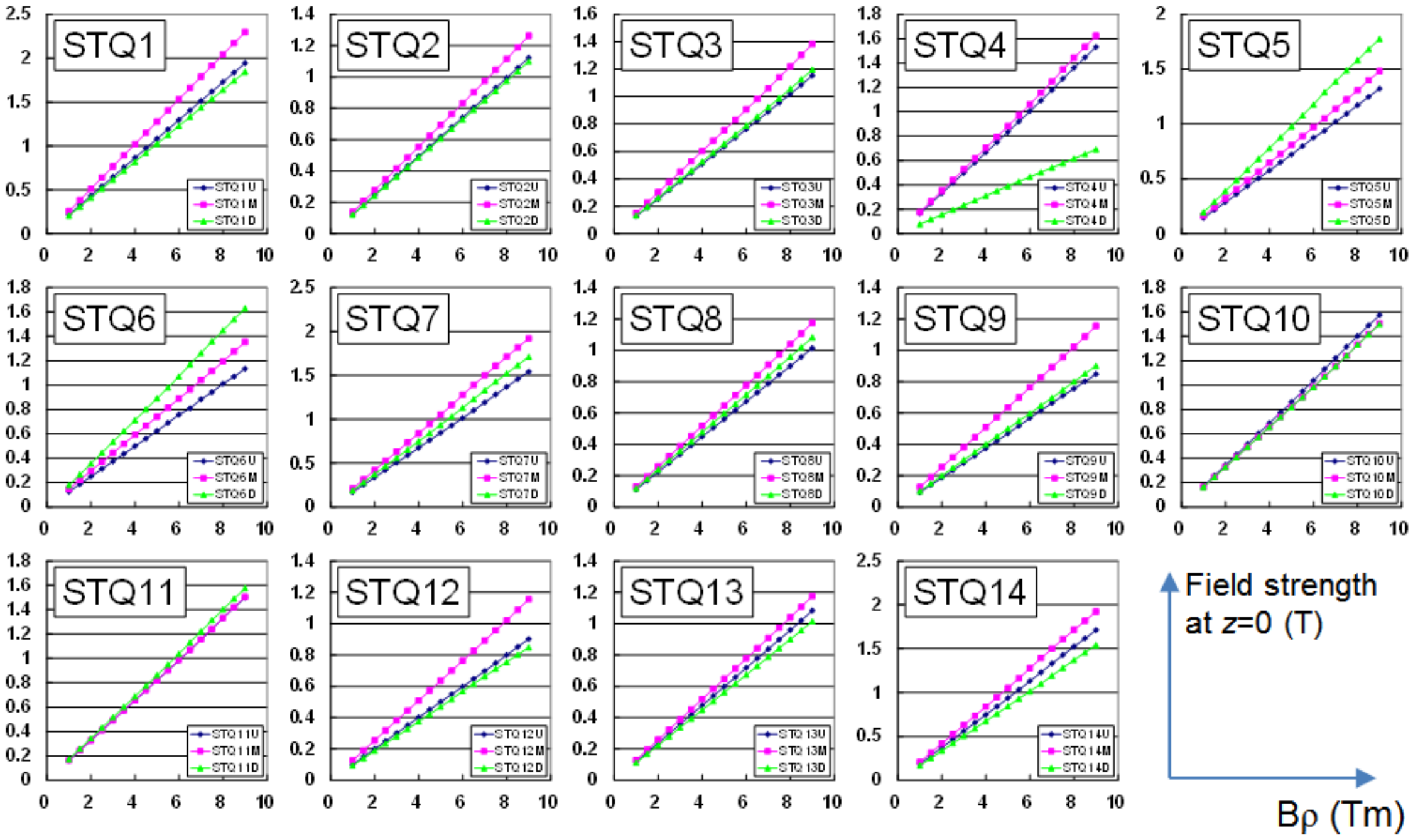

Fig. 13. Field strengths of the three quadrupoles in each STQ which are searched to satisfy the required ion-optical conditions in the BigRIPS standard mode. The field strengths at the warm bore radius of 120 mm and at the magnet center ($z$ = 0 mm) are plotted as functions of $\boldsymbol{B\rho}$ in each panel. The ion-optical search was made by the COSY calculations with the varying fringing fields in the FR2 mode. The blue, purple and green lines represent those of the upstream, middle and downstream quadrupoles in the STQ, respectively. For all quadrupoles, the field strengths are fitted by linear functions of the $\boldsymbol{B\rho}$ value. The resulting slope and offset parameters are stored in a configuration file of the magnet control system. See text.

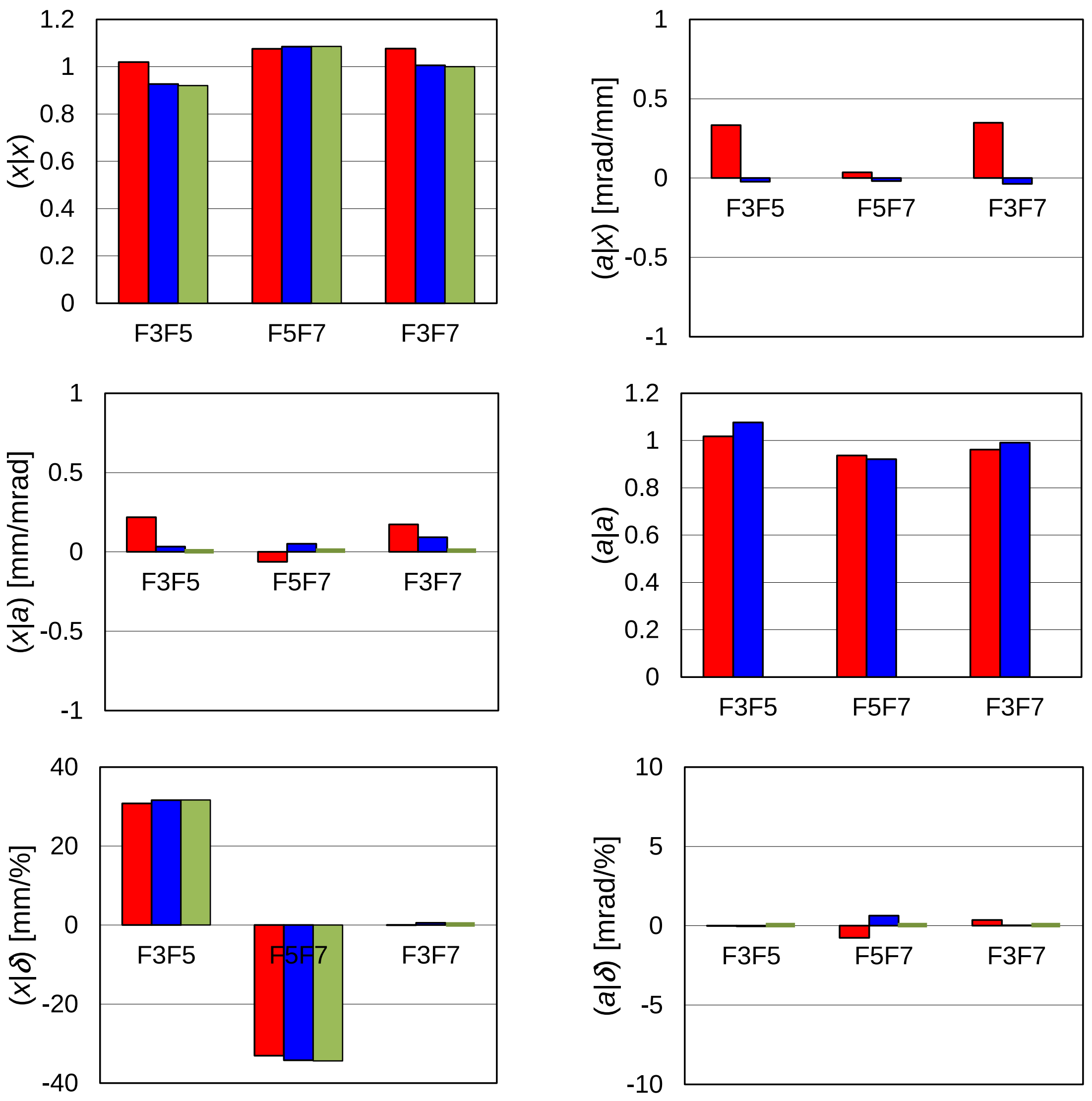

Fig. 14. The measured transfer matrix elements of F3-F5, F5-F7 and F3-F7 sections are compared with the COSY calculations. F3, F5 and F7 indicate the foci in the BigRIPS separator (see Refs. [11-13,15] for the details). The red and blue bars indicate the measured ones and the COSY results, respectively. The values used in the ion-optical search as the fitting conditions are also shown by the green bars. Note that $(\boldsymbol{a}|\boldsymbol{x})$ and $(\boldsymbol{a}|\boldsymbol{a})$ were not used as the fitting conditions.